\documentclass[fleqn,usenatbib,useAMS]{mnras}

\usepackage{graphicx}	% Including figure files
\graphicspath{{figs/}}  % figure path
\usepackage{amsmath}	% Advanced maths commands
\usepackage{amssymb}	% Extra maths symbols
\usepackage{multicol}        % Multi-column entries in tables
\usepackage{bm}		% Bold maths symbols, including upright Greek
\usepackage{pdflscape}	% Landscape pages
\usepackage[dvipsnames]{xcolor}
\usepackage{ulem}
\newcommand{\kms}{\ensuremath{{\rm km\ s^{-1}}}}		
\usepackage[T1]{fontenc}
\usepackage{ae,aecompl}
\usepackage{newtxtext,newtxmath}
\defcitealias{Zhu2023a}{Paper~I}
\defcitealias{Lu2023}{Paper~II}
\defcitealias{Zhu2023b}{Paper~III}
\defcitealias{Wang2023}{Paper~IV}
\defcitealias{Yang2007}{Yang07}
\usepackage{scalerel,tikz}
\usetikzlibrary{svg.path}
\definecolor{orcidlogocol}{HTML}{A6CE39}
\tikzset{orcidlogo/.pic={
		\fill[orcidlogocol] svg{M256,128c0,70.7-57.3,128-128,128C57.3,256,0,198.7,0,128C0,57.3,57.3,0,128,0C198.7,0,256,57.3,256,128z};
		\fill[white] svg{M86.3,186.2H70.9V79.1h15.4v48.4V186.2z}
		svg{M108.9,79.1h41.6c39.6,0,57,28.3,57,53.6c0,27.5-21.5,53.6-56.8,53.6h-41.8V79.1z M124.3,172.4h24.5c34.9,0,42.9-26.5,42.9-39.7c0-21.5-13.7-39.7-43.7-39.7h-23.7V172.4z}
		svg{M88.7,56.8c0,5.5-4.5,10.1-10.1,10.1c-5.6,0-10.1-4.6-10.1-10.1c0-5.6,4.5-10.1,10.1-10.1C84.2,46.7,88.7,51.3,88.7,56.8z};
}}
\newcommand\orcidicon[1]{\href{https://orcid.org/#1}{\mbox{\scalerel*{
				\begin{tikzpicture}[yscale=-1,transform shape]
					\pic{orcidlogo};
				\end{tikzpicture}
			}{|}}}}
\title[Density slopes of MaNGA galaxies]{MaNGA DynPop -- VI. Matter density slopes from dynamical models of 6000 galaxies versus cosmological simulations: the interplay between baryonic and dark matter}

\author[S. Li et al.]{
    Shubo Li\orcidicon{0009-0004-0904-7400}$^{1,2}$, 
    Ran Li\orcidicon{0000-0003-3899-0612}$^{1,3,2}$\thanks{E-mail: \url{ranl@bao.ac.cn}}, 
    Kai Zhu\orcidicon{0000-0002-2583-2669}$^{1,3,2,4}$, 
    Shengdong Lu\orcidicon{0000-0002-6726-9499}$^{4,5}$, 
    Michele Cappellari\orcidicon{0000-0002-1283-8420}$^{6}$, 
    Shude Mao\orcidicon{0000-0001-8317-2788}$^{4}$,\and
    \ Chunxiang Wang$^{1,3,2}$, 
    Liang Gao$^{1,3,2,5}$
    \\
    $^{1}$National Astronomical Observatories, Chinese Academy of Sciences, 20A Datun Road, Chaoyang District, Beijing 100101, China\\
    $^{2}$School of Astronomy and Space Science, University of Chinese Academy of Sciences, Beijing 100049, China\\
	$^{3}$Institute for Frontiers in Astronomy and Astrophysics, Beijing Normal University, Beijing 102206, China\\
    $^{4}$Department of Astronomy, Tsinghua University, Beijing 100084, China\\
    $^{5}$Institute for Computational Cosmology, Department of Physics, University of Durham, South Road, Durham, DH1 3LE, UK\\
    $^{6}$Sub-department of Astrophysics, Department of Physics, University of Oxford, Denys Wilkinson Building, Keble Road, Oxford, OX1 3RH, UK\\}

\date{Accepted 2024 March 18. Received 2024 March 11; in original form 2023 October 19}

\pubyear{2023}

\begin{document}
\label{firstpage}
\pagerange{\pageref{firstpage}--\pageref{lastpage}}
\maketitle

\begin{abstract}
%\michele{You had not yet updated the Abstract to reflect the changes in the paper. I rewrote it.}
We try to understand the trends in the mass density slopes as a function of galaxy properties. We use the results from the best Jeans Anisotropic Modelling (JAM) of the integral-field stellar kinematics for near 6000 galaxies from the MaNGA DynPop project, with stellar masses $10^9\ {\rm M_{\odot}}\la M_*\la 10^{12}\ {\rm M_{\odot}}$, including both early-type and late-type galaxies. We use the mass-weighted density slopes for the stellar $\overline{\gamma}_*$, dark $\overline{\gamma}_{_{\rm DM}}$ and total $\overline{\gamma}_{_{\rm T}}$ mass from the MaNGA DynPop project. As previously reported, $\overline{\gamma}_{_{\rm T}}$ approaches a constant value of $\overline{\gamma}_{_{\rm T}}\approx2.2$ for high $\sigma_{\rm e}$ galaxies, and flattens for $\lg(\sigma_{\rm e}/\kms)\la2.3$ galaxies, reaching $\overline{\gamma}_{_{\rm T}}\approx1.5$ for $\lg(\sigma_{\rm e}/\kms)\approx1.8$. We find that total and stellar slopes track each other tightly, with $\overline{\gamma}_{_{\rm T}}\approx\overline{\gamma}_*-0.174$ over the full $\sigma_{\rm e}$ range. This confirms the dominance of stellar matter within $R_{\rm e}$. We also show that there is no perfect conspiracy between baryonic and dark matter, as $\overline{\gamma}_*$ and $\overline{\gamma}_{_{\rm DM}}$ do not vary inversely within the $\sigma_{\rm e}$ range. We find that the central galaxies from TNG50 and TNG100 simulations do not reproduce the observed galaxy mass distribution, which we attribute to the overestimated dark matter fraction, possibly due to a constant IMF and excessive adiabatic contraction effects in the simulations. Finally, we present the stacked dark matter density profiles and show that they are slightly steeper than the pure dark matter simulation prediction of $\overline{\gamma}_{_{\rm DM}}\approx1$, suggesting moderate adiabatic contraction in the central region of galaxies. Our work demonstrates the power of stellar dynamics modelling for probing the interaction between stellar and dark matter and testing galaxy formation theories.
\end{abstract}

\begin{keywords}
galaxies: evolution – galaxies: formation – galaxies: kinematics and dynamics – galaxies: structure.
\end{keywords}

%%%%%%%%%%%%%%%%%%%%%%%%%%%%%%%%%%%%%%%%%%%%%%%%%%

%%%%%%%%%%%%%%%%% BODY OF PAPER %%%%%%%%%%%%%%%%%%

% The MNRAS class isn't designed to include a table of contents, but for this document one is useful.
% I therefore have to do some kludging to make it work without masses of blank space.
\begingroup
\let\clearpage\relax
%\tableofcontents
\endgroup
\newpage

\section{Introduction}

The prevailing Cold Dark Matter cosmological model proposes that structures in the Universe form hierarchically, where smaller dark matter halos form first and grow into larger ones through accretion and merger processes. If there were no baryonic matter, the evolution of the density profile of pure dark matter can be accurately traced using N-body numerical simulations. Various studies have shown that  in a universe of cold dark matter, the density of dark matter at the inner regions of halos changes with radius according to a $r^{-1}$ law \citep{NFW1996, NFW1997, Springel_Aquarius, Gao_Phoenix, WangJie2020}. 

However, in the real Universe, baryonic matter condenses at the centre of dark matter haloes, forming galaxies, and the galaxies co-evolve with the dark matter halo. Thus, the total material density profile at the center of a dark halo is the result of a combination of both baryonic and dark matter. Galaxies of different types possess varying stellar mass density profiles. For instance, early-type galaxies (ETGs) have stellar density slope with $\gamma_*\sim2.3$, assuming a mass-follows-light case, while the stellar density slope (slope of the luminosity profile) of late-type galaxies (LTGs) is noticeably flatter, with some galaxies even approaching $\gamma_*\sim1.2$ \citep{LiRan2019}. In general, it is clear that if the halo has density slope of dark matter $\gamma_{_{\rm DM}}$ and star $\gamma_*$, the total density slope $\gamma_{_{\rm T}}$ can vary between $\gamma_*$ and $\gamma_{_{\rm DM}}$ when the dark matter fraction $f_{\rm DM}$ varies between 0 and 1 in the same radial range. As a result, $\gamma_{_{\rm T}}$ of a galaxy can vary significantly depending on the ratio of dark matter to baryonic matter at the center of the dark halo.

Apart from this ratio, the interplay between baryonic matter and dark matter during the galaxy formation process can also alter the overall material density profile in the center of the dark halo. Firstly, the collapse and accumulation of baryonic matter in the center of the dark halo may induce an ‘adiabatic contraction’, making the distribution of dark matter more concentrated \citep{Blumenthal1986, Gnedin2004, Gustafsson2006}. Secondly, the formation and subsequent evolution of stars can potentially impact the structure of the dark matter halo. For example, the feedback effects of stars and supermassive black holes can swiftly expel gas from the central region, thereby changing the gravitational potential well of the dark halo, resulting in a flatter central density profile of the dark halo \citep[e.g.][]{Read2005, Pontzen2012, Bentez2018, Bose2019}.

Deviations from ‘cold’ dark matter properties can also cause differences in the density profiles of dark matter halos' inner regions. In self-interacting dark matter model, scattering among the dark matter particles can generate a core-like profile in the centre of dark matter halo \citep{Kaplinghat2016, Tulin2018, Robertson2019, Bondarenko2021, Andrade2022, Eckert2022}.

Observationally, density profiles can be measured using dynamics and gravitational lensing. For spiral galaxies, the rotation curves of the gas disk can provide an excellent probe of galaxy mass distribution \citep[e.g.][]{Bosma1978, Faber1979, Rubin1980}. Recently, \citet{Tortora2019} analyzed the rotation curves of 175 spiral galaxies from the SPARC project \citep{Lelli2016}, calculating the mass-weighted density slopes for these galaxies. They found that the density profiles of these galaxies become steeper as galaxy mass increases. At the low-mass end (less than $10^{8}\ {\rm M_{\odot}}$), the total matter density profile slope of the galaxies is nearly 1, while at the high-mass end (greater than $10^{10.5}\ {\rm M_{\odot}}$), the density profile slope of the galaxies is close to 2, indicating an isothermal density profile.

Galaxy-galaxy strong gravitational lensing is typically applied to ETGs of high velocity dispersion. The Sloan Lens Advanced Camera for Surveys \citep[SLACS;][]{SLACS} searched for lensing candidates within the SDSS survey, and through subsequent observations with the Hubble Space Telescope, high-resolution images were obtained for lens modeling. For galaxies with stellar masses greater than $10^{11}\ {\rm M_{\odot}}$, they discovered that the average inner density profile slope of total mass $\langle\gamma_{_{\rm T}}\rangle=2.078$ \citep{Koopmans2009, Auger2010}, which is very close to predictions made by the isothermal model. This result has been confirmed by subsequent similar strong gravitational lensing analyses \citep{Sonnenfeld2013, LiRui2018, Etherington2023} and the near-isothermal feature of ETGs is sometimes termed as the ‘bulge-halo conspiracy’ (see Section~\ref{Sec4.2} for details).

Compared to strong gravitational lensing and gas dynamics, stellar kinematics can be more widely applied to different types of galaxies over a broad range of galaxy masses. \citet{Cappellari2015} was the first to systematically apply the Jeans Anisotropic Modelling (JAM) method to analyze the dynamic data of 14 large-mass ETGs, reporting a nearly ‘universal’ average total density profile slope of $\langle \gamma_{_{\rm T}} \rangle=2.19$, with small scatter. Using the JAM method, \citet{Poci2017} analyzed 260 ETGs from the ATLAS$^{\rm 3D}$ survey \citep{Cappellari2013}. They found for galaxies with velocity dispersion higher than a threshold $\lg(\sigma_{\rm e}[\kms]) \sim 2.1$, $\langle \gamma_{_{\rm T}} \rangle=2.193$, but for low velocity dispersion galaxies, $\gamma_{_{\rm T}}$ shows a decreasing trend with decreasing velocity dispersion.

Thanks to the SDSS-IV MaNGA survey \citep{Bundy2015}, a new generation of Integral Field Unit (IFU) survey, a broad range of galaxy types over a wide mass range have obtained stellar kinematic data usable for dynamic modeling. \citet{LiRan2019} analyzed MaNGA galaxies in SDSS Data Release 14 \citep[SDSS DR14;][]{Abolfathi2018}, which covers a mass range from $10^9\ {\rm M_{\odot}}$ to $10^{12}\ {\rm M_{\odot}}$, including both early- and late-type galaxies. Using the mass distribution models obtained with the JAM method from \citet{Li2018}, they computed the total density slope of these galaxies. For ETGs with $\sigma_{\rm e}>100\ \kms$, they obtained similar results to previous studies, with the average mass-weighted total density slope $\langle\overline{\gamma}_{_{\rm T}}\rangle=2.24$. Moreover, they showed not only for ETGs, but for all types of galaxies as a whole, how the total density slope decreases with decreasing velocity dispersion below a velocity dispersion threshold. Additionally, they showed that central galaxies in clusters have a flatter $\overline{\gamma}_{_{\rm T}}$ than satellite galaxies.

In this study, we utilize the mass distribution models derived from the MaNGA DynPop project \citep{Zhu2023a, Zhu2023b, Lu2023, Lu2023b, Wang2023}, which analyze the full data release of MaNGA survey that contains the IFU data for over 10,000 nearby galaxies. This dataset cover a wide range of galaxy types and stellar masses, from $10^9$ to $10^{12}\ {\rm M_{\odot}}$ \citep{Wake2017}. The MaNGA DynPop project takes advantage of this IFU data for galaxy dynamical modeling \citep[][Paper I, hereafter]{Zhu2023a}. In \citet[][Paper II, hereafter]{Lu2023}, we obtain the stellar population properties of MaNGA galaxies under the Salpeter \citep{Salpeter} initial mass function (IMF) assumption, including the stellar mass-to-light ratio used here to calculate the stellar mass $M_*$. In \citet[][Paper III, hereafter]{Zhu2023b}, we calculate the total density slope for 6000 MaNGA galaxies. Our results once again confirm that that for ETGs, the average mass-weighted total density slope is slightly steeper (approximately 2.2) than the isothermal case. This finding is consistent with the previously observed trend in the variation of galaxy total density slopes with velocity dispersion, as reported by \citet{LiRan2019}. In \citet[][Paper~IV, hereafter]{Wang2023}, we select central galaxies of clusters and groups that have weak gravitational lensing measurements, and jointly constrain their density profile slopes using both dynamics and weak gravitational lensing. We show that the inner total density profiles of central galaxies in clusters and groups are close to isothermal profile, and find that such density profiles require the stellar mass of the galaxies to be significantly higher than the stellar mass derived from stellar population synthesis models assuming a Chabrier \citep{Chabrier} IMF.

In this paper, we will use the galaxy mass distribution models obtained from JAM to more comprehensively analyze the evolution of galaxy density slope with galaxy properties, the connection between the slopes of total and stellar density, along with the dark matter fraction, and the comparison of observational results with hydrodynamic numerical simulations. Additionally, we will also analyze the stacked density profile of galaxies.

The structure of this paper is as follows. In Section~\ref{data} , we briefly introduce the MaNGA project and the simulation data we use for comparison. Section~\ref{method} provides an brief overview of the dynamical modeling methods used in MaNGA DynPop, as well as the method we use to compute the mass-weighted density slope in this paper. Our results are presented in Section~\ref{result}. In Section~\ref{discussion}, based on the results of this study, we discuss the bulge-halo conspiracy scenario, the implications of our comparison with simulations, and the impact of the quality flag in our results. The final section, Section~\ref{summary}, summarizes our conclusions.

Throughout this paper, we follow \citetalias{Zhu2023a} and adopt the Planck ${\rm \Lambda CDM}$ cosmology \citep{Planck2015} with $\Omega_m = 0.307$ and $H_0 = 67.7\ {\rm km}\cdot{\rm s}^{-1}\cdot{\rm Mpc}^{-1}$.

\section{Data}
\label{data}
\subsection{MaNGA galaxies}
\label{sec:MaNGA}
This paper investigates density slopes of galaxies in the final data release of the MaNGA survey \citep[SDSS DR17;][]{Abdurrouf2022}, which includes unprecedentedly more than 10000 nearby galaxies. For further information of the MaNGA survey, the readers are referred to papers listed below: an overview of MaNGA \citep{Bundy2015}, SDSS-IV technical summary \citep{Blanton2017}, MaNGA instrumentation \citep{Drory2015}, sample design \citep{Wake2017}, observing
strategy \citep{Law2015}, spectrophotometric calibration \citep{Smee2013, Yan2016a}, survey execution and initial data quality \citep{Yan2016b} and an overview of SDSS telescope is included in \citet{Gunn2006}.

We adopt the mass-weighted density slopes (see Section~\ref{density slope} for details) derived in \citetalias{Zhu2023a}, which construct accurate mass models using the Jeans Anisotropic Modelling \citep[JAM;][]{Cappellari2008, Cappellari2020} method and the MaNGA stellar kinematics extracted from the Data Analysis Pipeline \citep[DAP;][]{Belfiore2019, Westfall2019}. A brief introduction to the dynamical modelling is presented in Section~\ref{sec:JAM}. We select galaxies with an acceptable visual modelling quality (i.e. $\rm Qual \geqslant 1$) to ensure that the density slopes are reliable. For these approximately 6000 galaxies, we require the difference in the mass-weighted total density slope between $\rm JAM_{cyl}$ and $\rm JAM_{sph}$ modelling (see Section~\ref{sec:JAM}) is smaller than three times of 0.079, which is the observed root-mean-square (rms) scatter of this dynamical property among different model assumptions for galaxies with $\rm Qual=1$ (see Fig. 13 and Table 3 of \citetalias{Zhu2023a}), to ensure the reliability of our conclusion. In result, our final MaNGA sample includes 5688 galaxies. For more information of modelling qualities and the dynamical modeling, readers are referred to \citetalias{Zhu2023a}.

\subsection{IllustrisTNG galaxies}
We compare the measured mass-weighted density slopes with theoretical prediction derived from The Next Generation Illustris Simulations (IllustrisTNG, TNG hereafter), which is a suite of state-of-the-art magneto-hydrodynamic cosmological galaxy formation simulations carried out in large cosmological volumes with the moving-mesh code \textsc{arepo} \citep{arepo}. The TNG collaboration has publicly released the general properties of galaxies, which are available for use \citep{TNG-release}. In this paper, we use the highest resolution version of two simulations from the suite: the TNG 50 \citep{TNG50-1, TNG50-2}, which is a full-physics version with a box size of 51.7 Mpc and has a mass resolutions for both baryonic and dark matter component of $m_{\rm baryon} = 8.5 \times 10^5\ {\rm M_\odot}$ and $m_{\rm DM} = 4.5 \times 10^5\ {\rm M_\odot}$, respectively; the TNG100 \citep{TNG100-1, TNG100-2, TNG100-3, TNG100-4, TNG100-5}, which is a full-physics version with a cubic box of 110.7 Mpc side length and has a mass resolution for baryonic and dark matter component of $m_{\rm baryon} = 1.4 \times 10^6\ {\rm M_\odot}$ and $m_{\rm DM} = 7.5 \times 10^6\ {\rm M_\odot}$, respectively. The gravitational softening length for dark matter and stellar particles is $\epsilon_{\rm softening} = 0.288$ kpc for TNG50 and $\epsilon_{\rm softening} = 0.738$ kpc for TNG100. Galaxies residing within their host dark matter halos are identified using the \textsc{subfind} algorithm \citep{subfind-1, subfind-2}. To match the MaNGA observation, we select central TNG galaxies from snapshot 99 (corresponding to redshift $z=0$) with stellar mass $M_* \geqslant 10^9\ {\rm M_\odot}$. Combined with the selection criterion based on galaxy size (see Section~\ref{density slope}), our final TNG sample ended up with 1733 TNG50 galaxies and 6107 TNG100 galaxies.

\section{Method}
\label{method}
\subsection{Dynamical modelling}
\label{sec:JAM}
In \citetalias{Zhu2023a}, we use the axisymmetric Jeans Anisotropic Modelling \citep[JAM;][]{Cappellari2008,Cappellari2020} method to derive the dynamical quantities for the whole MaNGA sample and provide the values of each quantity for eight models. Among the eight models, we adopt two assumptions of velocity ellipsoids (cylindrically-aligned model $\rm JAM_{cyl}$ and spherically-aligned model $\rm JAM_{sph}$) and four mass models (for each velocity ellipsoid assumption) that have different assumptions on the dark matter distribution. The surface brightness of each galaxy is obtained from Multi-Gaussian Expansion \citep[MGE;][]{Emsellem1994,Cappellari2002} fitting to the SDSS $r-$band images, and then is deprojected to obtain the luminosity distribution of the kinematic tracers. The total mass distribution (or equivalently the total gravitational potential) is described as the luminosity density multiplied by a spatially constant stellar mass-to-light ratio and the dark matter mass distribution. For a given gravitational potential, the modelled velocity second moments derived from the axisymmetric Jeans equations are compared to the observed one to determine the best-fitting parameters. The JAM method has been tested on mock galaxies generated using cosmological hydrodynamical simulation, which has shown that the density slopes of galaxies can be recovered robustly \citep{LHY2016}. 

We mainly use the $\rm JAM_{cyl}$ + generalized Navarro-Frenk-White (gNFW) dark model in this work, which is the most flexible mass model provided in \citetalias{Zhu2023a} and is supposed to provide the most accurate measurements of density slopes. The gNFW profile \citep{Wyithe2001} is described as 
\begin{equation}
    \rho_{_{\rm DM}}(r) = \rho_s\left(\frac{r}{r_s}\right)^{-\gamma}\left(\frac{1}{2}+\frac{1}{2}\frac{r}{r_s}\right)^{\gamma-3},
\end{equation}
where $r_s$ is the characteristic radius, $\rho_s$ is the characteristic density, and $\gamma$ is the inner density slope. For $\gamma=1$, this function reduces to the NFW profile. The density slopes of other mass models can be used to access the systematic uncertainties, which are listed in Table 3 of \citetalias{Zhu2023a}. Specifically, for our sample with $\rm Qual\geqslant1$, the rms scatter of the mass-weighted total density slopes between different models is smaller than 0.079, indicating the robustness of density slopes measurements. We adopt three times of this value in Section~\ref{sec:MaNGA} to filter galaxies as a guarantee of reliability in JAM modeling. In the following sections, unless explicitly stated otherwise, all the dynamical properties of MaNGA galaxies utilized are from the catalogue of \citetalias{Zhu2023a}.  

\subsection{Mass-weighted density slope}
\label{density slope}
With galaxy density profile approximated by power-law $\rho \propto r^{-\gamma}$, one can follow \citet{Dutton2014} to calculate the mass-weighted density slope within the effective radius $R_{\rm e}$ as
\begin{align}
\overline{\gamma} \equiv -\frac{1}{M(r<R_{\rm e})}\int_{0}^{R_{\rm e}}4\pi r^2\rho (r)\frac{\mathrm{d}\log\rho}{\mathrm{d}\log r}\mathrm{d}r = 3-\frac{4\pi R_{\rm e}^3\rho (R_{\rm e})}{M(r<R_{\rm e})}.
\label{eq: MWslope}
\end{align}
For MaNGA galaxies, the projected circularized half-light radius $R_{\rm e}$ (‘Re\_arcsec\_MGE’ in the catalogue of \citetalias{Zhu2023a}) is derived from MGE fitting of MaNGA galaxy's r-band image. For TNG galaxies, we did not directly use the 3D half-mass radius provided in the TNG subhalo catalog but assumed that the projection direction was along the x-axis and based on the surface mass density of the stellar particles in the y-z plane, we calculated the projected radius of a circle enclosing half of the galaxy stellar particles as $R_{\rm e}$. Throughout this paper, we use $\overline{\gamma}_{\rm x}$ to denote the mass-weighted density slope, where x represents the material components we are considering. We use ‘T’, ‘*’ and ‘DM’ to label total, stellar and dark matter, respectively.

The mass-weighted density slopes of MaNGA galaxies come from the catalogue of \citetalias{Zhu2023a}. For TNG galaxies, we divide galaxy's radius from $\epsilon_{\rm softening}$ to 100 kpc into 50 equal intervals in logarithmic space to obtain the density and enclosed mass profile and interpolate them as function of radius. When considering the total mass, we combine the masses of star, dark matter, and gas particles while when focusing on the mass of dark matter, we sum up the masses of dark matter and gas particles. In cosmological N-body simulations, the gravitational softening length $\epsilon_{\rm softening}$ is introduced into the Newtonian gravity formula to prevent close encounters between particles. In order to mitigate the impact of unrealistic gravitational softening on the density profiles of simulated galaxies \citep[see details in][]{Softening}, it is necessary to exclude the innermost region of the simulated galaxies. Similar to \citet{WangTNG1}, we set $0.3\ R_{\rm e}$ as the lower integral limit \citep[in][they choose $0.4\ R_{\rm e}$ to include more ETG samples]{WangTNG1} and only choose simulated galaxies with $0.3\ R_{\rm e} > \epsilon_{\rm softening}$. We should aslo change Eq.~\ref{eq: MWslope} into
\begin{align}
\overline{\gamma} &\equiv -\frac{1}{M(r_{\rm min}<r<R_{\rm e})}\int_{r_{\rm min}}^{R_{\rm e}}4\pi r^2\rho (r)\frac{\mathrm{d}\log\rho}{\mathrm{d}\log r}\mathrm{d}r\notag\\
&= 3-\frac{4\pi R_{\rm e}^3\rho (R_{\rm e})}{M(r_{\rm min}<r<R_{\rm e})}+\frac{4\pi r_{\rm min}^3\rho (r_{\rm min})}{M(r_{\rm min}<r<R_{\rm e})}
\end{align}
and substitute $0.3\ R_{\rm e}$ into $r_{\rm min}$. For TNG50, 92 per cent (1733) central galaxies satisfy this criterion. But for TNG100, due to the larger $\epsilon_{\rm softening}$, only 50 per cent (6107) central galaxies satisfy this criterion. However, we have verified (although we do not show) that the results presented in the remainder of this paper remain unchanged, regardless of the inclusion or exclusion of these TNG100 galaxies with smaller size. Furthermore, if we change the filtering criteria to $0.3\ R_{\rm e} > 2.5\epsilon_{\rm softening}$, it will not affect our conclusion.

\section{Results}
\label{result}
\subsection{Total density slope}
\begin{figure*}
        \centering
        \includegraphics[width=2\columnwidth]{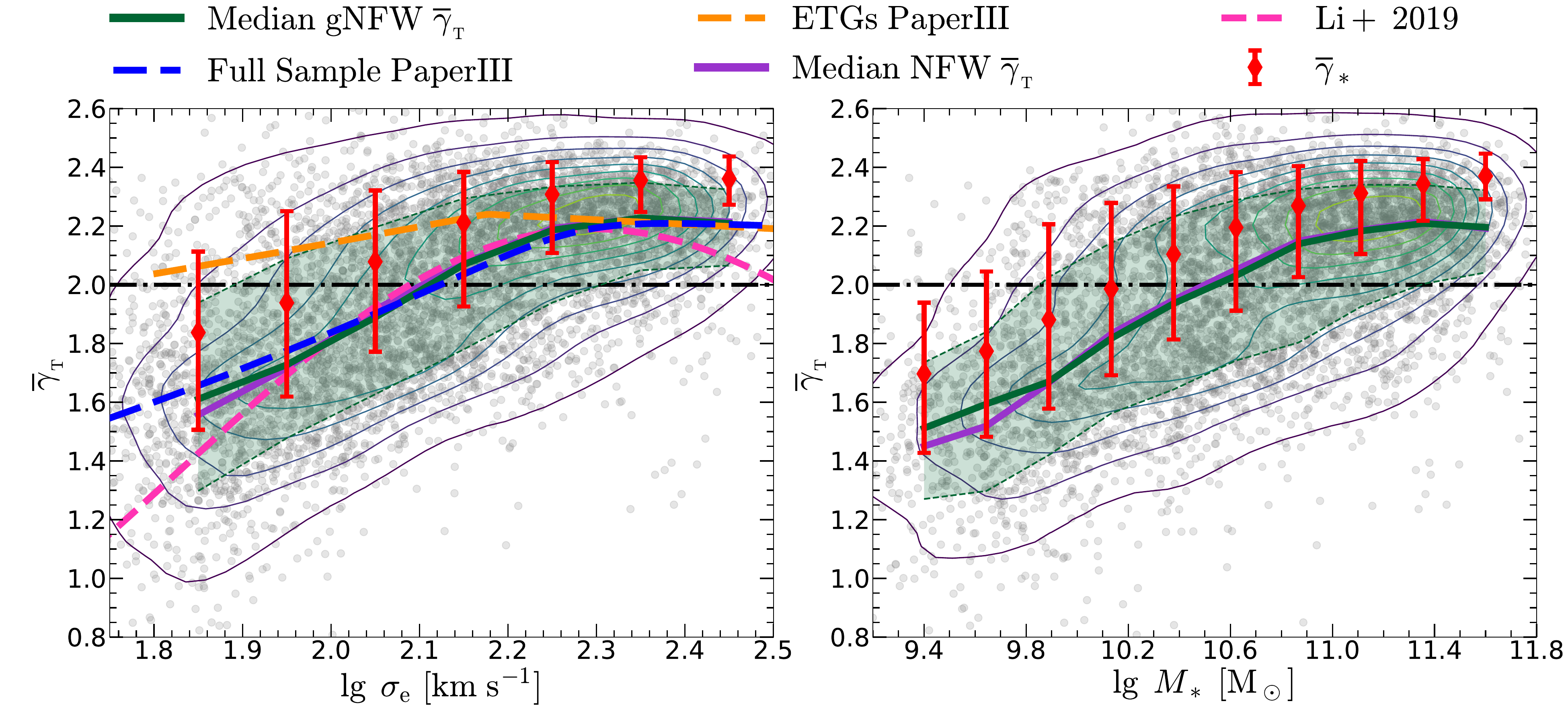}
        \caption{%\michele{(1) The lines at low sigma show a sharp bend (especially the red line). However, you have very few galaxies in the last few sigma bins. This seems a spurious artifact, due to few galaxies and incompleteness. How are you doing the computation? Are you considering the same number of galaxies in each sigma bin? If not, you should try that. Or alternatively, you could stop the median curve at larger sigma, excluding from the median e.g. galaxies with lgSigma<1.8.}
        $\overline{\gamma}_{_{\rm T}}$ as a function of $\sigma_{\rm e}$ (left) and $M_*$ (right). Grey scattered points represent $\overline{\gamma}_{_{\rm T}}$ derived under the ${\rm JAM{cyl}}$ + gNFW assumption. The solid green line represents the median of $\overline{\gamma}_{_{\rm T}}$, with the green shaded region denotes the 1-$\sigma$ scatter. 
        % \michele{(2) The error of the median would be useful if we were measuring the slope of a single galaxy by repeated observations. But here, we have a distribution of values coming from different galaxies, which have intrinsically different slopes. More meaningful are the [16th, 84th] percentiles, as you show in Fig.2,3. These give an idea of the spread of the distribution. You should *always* show percentiles, here and also in figs. 6, 7, 11, 12, 13. In fact, you had percentiles in a previous version of the figs.6 ,7, but you removed them.}. 
        The medians of NFW $\overline{\gamma}_{_{\rm T}}$ are also plotted. The fitting result of $\overline{\gamma}_{_{\rm T}}$-$\sigma_{\rm e}$ relations for the full sample and ETGs in \citetalias{Zhu2023b} are represented by blue and orange dashed lines, respectively.
        %\michele{(3) Please use labels like ‘Full Sample Paper III’ and ‘ETGs Paper III’ instead of ‘fitting’, to make clear they come from another paper, as you did for ‘Li+2019’. For the stellar density, it is a bit confusing to write gNFW. The density only comes from the stars, and there are minimal differences due to the fitted inclination. I would just write ‘Median $\gamma_*$’ to avoid confusion. And write in the text that the inclination comes from the gNFW models.} 
        The result of \citet{LiRan2019} is plotted by pink dashed line. The median 
        and the 1-$\sigma$ scatter are showed by red diamonds with error bars. The black contours show the kernel density estimate for the sample distribution. The slope value of the isothermal profile is depicted with a black horizontal dash-dot line.}
        \label{scaling}
    \end{figure*}

 In Fig.~\ref{scaling}, we present the scaling relations between the mass-weighted total density slopes $\overline{\gamma}_{_{\rm T}}$ and the velocity dispersion within effective radius $\sigma_{\rm e}$ or the stellar mass $M_*$. $\sigma_{\rm e}$ (‘Sigma\_Re’ in the catalogue of \citetalias{Zhu2023a}) is defined as the square root of luminosity-weighted second velocity moments within the elliptical half-light isophote (‘Rmaj\_arcsec\_MGE’ in the catalogue of \citetalias{Zhu2023a}) and is calculated as
\begin{equation}
\sigma_{\rm e} \approx \langle v_{\rm rms}^2\rangle_{\rm e}^{1/2} = \sqrt{\frac{\Sigma_k F_k (V_k^2+\sigma_k^2)}{\Sigma_k F_k}},
\end{equation}
where $F_k$, $V_k$and $\sigma_k$ are the flux, stellar velocity, and stellar velocity dispersion in the $k$-th IFU spaxel.
For each galaxy, $M_*$ of Salpeter IMF is calculated as
\begin{equation}
M_* = (M_*/L)_{\rm SPS}(<R_{\rm e})\times L,
\end{equation}
where $(M_*/L)_{\rm SPS}(<R_{\rm e})$ (comes from ‘ML\_int\_Re’ in the catalogue of \citetalias{Lu2023}) is the r-band stellar mass-to-light ratio obtained from the stacked spectrum again within the elliptical half-light isophote, while $L$ (comes from ‘Lum\_tot\_MGE’ in the catalogue of \citetalias{Zhu2023a}) represents the r-band total luminosity derived from MGE models. To compare with the simulations which assume Chabrier IMF, we then subtract the $M_*$ by 0.25 dex \citep{offset}. As demonstrated in Figure 8 of \citetalias{Zhu2023b}, $\overline{\gamma}_{_{\rm T}}$ increases rapidly with velocity dispersion from 1.6 to 2.2 for galaxies with $\lg(\sigma_{\rm e}[\kms]) \la  2.28$ ($\approx 190\ \kms$). Above this velocity dispersion, the median $\overline{\gamma}_{_{\rm T}}$ remain relatively constant at 2.2, slightly steeper than the slope of isothermal profile (that is, 2). This trend can be well described by equation (13) of \citetalias{Zhu2023b}, which reads:
\begin{equation}
\label{eq: fitting}
\overline{\gamma}_{_{\rm T}} = A_{0}\left(\frac{\sigma_{\rm e}}{\sigma_b}\right)^\gamma\left[\frac{1}{2}+\frac{1}{2}\times\left(\frac{\sigma_{\rm e}}{\sigma_b}\right)^\alpha\right]^\frac{\beta-\gamma}{\alpha},
\end{equation}
with $\{A_0, \sigma_b, \alpha, \beta, \gamma\} =\{2.18, 189, 11.13, -0.02, 0.34\}$ the best-fitting parameters. $\overline{\gamma}_{_{\rm T}}$ increases with $M_*$ as well. A mild transition can be found at $M_* \approx 10^{11}\ M_\odot$, above which the median value of $\overline{\gamma}_{_{\rm T}}$ is constantly 2.2.

Here we make a comparison with the results of \citet{LiRan2019}. While the overall trend of the $\overline{\gamma}_{_{\rm T}}$–$\sigma_{\rm e}$ relation is similar to the findings in \citet{LiRan2019}, the transition of the slope decrease occurs at a lower value of $\sigma_{\rm e}$ in \citet{LiRan2019}, at around 100 $\kms$, and the trend difference is less apparent in the plot because logarithmic coordinates are used. We infer this discrepancy is mainly due to the difference of measurements in $\sigma_{\rm e}$ for different data releases. With the updated MaNGA DAP \citep{Law2021}, it can be found that the velocity dispersion in SDSS DR17 had been underestimated in previous MaNGA data releases, especially for low-mass galaxies. Therefore, the same galaxy at the final data release will have a higher velocity dispersion and in turn have a steeper total density profile predicted by dynamical modelling. As a result, the transition of the $\overline{\gamma}_{_{\rm T}}$–$\sigma_{\rm e}$ relation is smoother than what is observed in SDSS-IV DR14. 

Based on the MaNGA Deep Learning morphological catalogue \citep{MDL} and adopt the most restrictive selection strategy listed in sec. 3.4.1 of \citet{MDL}, we select 2180 ETGs from our MaNGA samples. The best-fitting equation for the $\overline{\gamma}_{_{\rm T}}$–$\sigma_{\rm e}$ relation in ETGs can be found in Table 1 of \citetalias{Zhu2023b}, and it shares the same form as the one for the full sample. The parameter values for this equation are $\{A_0, \sigma_b, \alpha, \beta, \gamma\} = \{2.24, 150, 397.85, -0.03, 0.11\}$. The turnover occurs earlier at $\sigma_{\rm e}\approx150\ \kms$. The gradient of the decreasing slope with decreasing velocity dispersion becomes gentler, reflecting the fact that ETGs generally exhibit steeper profiles. 

We also plot the mass-weighted stellar density slope $\overline{\gamma}_*$ derived under the gNFW model in Fig.~\ref{scaling}. Interestingly, the trend of the mass-weighted density slope of the stellar component is very similar to that of the total mass. We fit the $\overline{\gamma}_*-\sigma_{\rm e}$ relation using a function of the same form as for $\overline{\gamma}_{_{\rm T}}$ with $\{A_0, \sigma_b, \alpha, \beta, \gamma\} =\{2.33, 190, 9.98, 0.03, 0.25\}$ the best-fitting parameters. The mean difference between the fitted values of $\overline{\gamma}_*$ and $\overline{\gamma}_{_{\rm T}}$ is consistently 0.174 over the full $\sigma_{\rm e}$ range with the standard deviation of the differences is roughly 0.016, indicating that the total and stellar slopes closely follow a similar trend. As shown in Section 3.4 of \citetalias{Zhu2023b} and Fig.~\ref{fdm_Ms}, the JAM modelling generally predicts galaxies with low dark matter fraction within $R_{\rm e}$, $f_{\rm DM}(<R_{\rm e})$. Especially for ETGs, 90 per cent of them have $f_{\rm DM}(<R_{\rm e})<23$ per cent. Therefore, the similarity of $\overline{\gamma}_*$ and $\overline{\gamma}_{_{\rm T}}$ may be a direct result of the stellar component dominating the total mass budget within the effective radius.

From Fig.~\ref{scaling} we can see that gNFW and NFW model assumption derive almost identical median scaling relations on $\overline{\gamma}_{_{\rm T}}$ (except for small $\sigma_{\rm e}$ and $M_*$ end), reflecting the reliability of JAM modelling. In the following figures, unless otherwise specified, we all display the results of MaNGA galaxies under ${\rm JAM_{cyl}}$ + gNFW model. 

\begin{figure*}
        \centering
        \includegraphics[width=2\columnwidth]{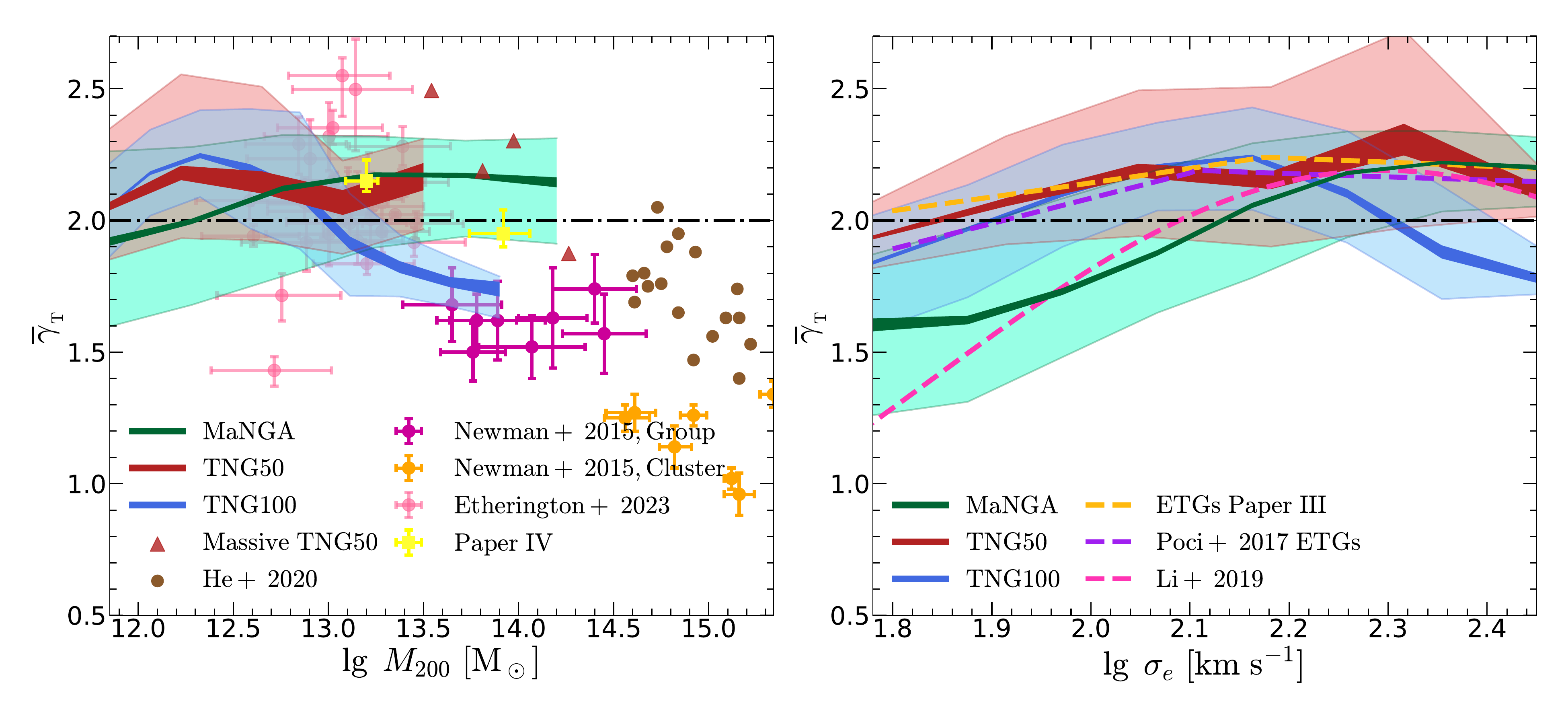}
        \caption{ $\overline{\gamma}_{_{\rm T}}$ as a function of $M_{200}$ (Left) and $\sigma_{\rm e}$ (right) for central galaxies. The median of MaNGA, TNG50, and TNG100 is represented by green, red, and blue lines, respectively. The line width corresponds to the standard error of the median. The shaded region in the same color represents the 1-$\sigma$ scatter. The four most massive galaxy of TNG50 are shown by red solid triangles in left panels. The data points with error bars in left panel represent the observational results from \citet{Etherington2023} for SLACS galaxies, \citet{Newman2015} for central galaxies of group- and cluster-scale and \citetalias{Wang2023} for central MaNGA galaxies of group- and cluster-scale combining JAM with weak lensing. The results from the C-EAGLE simulation \citep{He2020} are also included. The dashed lines in the right panel represent the best-fitting results from \citet{Poci2017} and \citet{LiRan2019}, while the result for ETGs is from \citetalias{Zhu2023b}. The slope value of the isothermal profile is shown with a black horizontal dash-dot line.}
        \label{Compare}
    \end{figure*}

In Fig.~\ref{Compare}, we present $\overline{\gamma}_{_{\rm T}}$ as a function of halo mass $M_{\rm 200}$ (left) and $\sigma_{\rm e}$ (right) for central galaxies. We adopt the central-satellite classification results and $M_{200}$  from an updated version\footnote{\url{https://gax.sjtu.edu.cn/data/Group.html}} of the SDSS DR4 group catalog \citep{Yang2007} (SDSSGC, hereafter). Following SDSSGC, there are 3831 central galaxies (the galaxy with the largest stellar mass in one group) in our MaNGA samples, of which 2922 have halo mass that are considered reliable and are plotted in left panel of Fig.~\ref{Compare}. The median $\overline{\gamma}_{_{\rm T}}$ exhibits a subtle increase from 1.9 to about 2.2 among central galaxies with $M_{200} \la 10^{13}\ {\rm M_\odot}$, and remains consistently fixed at 2.17 when $M_{200} \ga 10^{13.5}\ {\rm M_\odot}$. The behavior of $\overline{\gamma}_{_{\rm T}}$-$\sigma_{\rm e}$ for central galaxies is the same as that for full sample, except this relationship has a slightly more gentle turn for central galaxies. \citet{LiRan2019} found that, for different values of $\sigma_{\rm e}$, the mean $\overline{\gamma}_{_{\rm T}}$ of satellite galaxies is overall about 0.1 higher than that of central galaxies, suggesting that this may be due to differences in the galactic formation background. Here, we further confirm this phenomenon with SDSS DR17.

\begin{figure*}
    \centering
    \includegraphics[width=2\columnwidth]{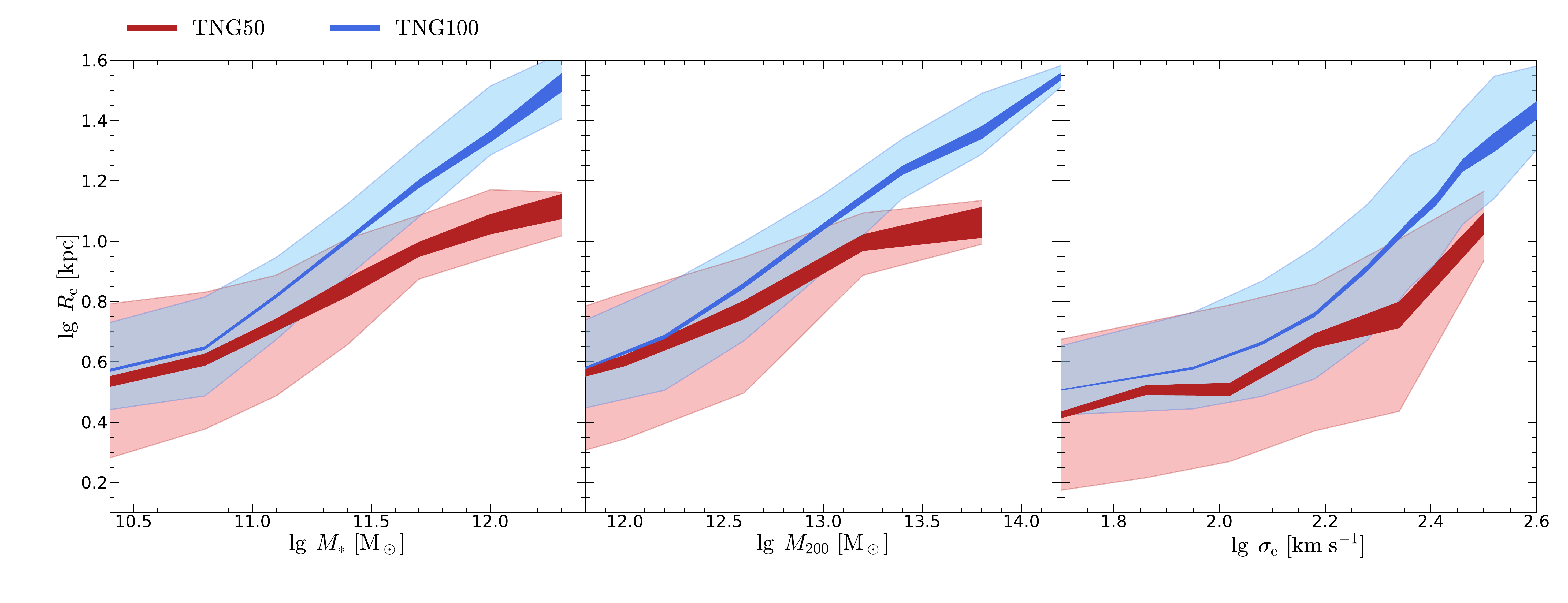}
    \caption{ Compare the $R_{\rm e}-M_{*}$ (left), $R_{\rm e}-M_{200}$ (middle) and $R_{\rm e}-\sigma_{\rm e}$ (right) relation for central galaxies between TNG50 and TNG100. The solid lines shows their median with the line width corresponding to the standard error of the median. The shadow regions indicate the 1-$\sigma$ scatters.}
    \label{50vs100}
\end{figure*}

We also compare the results of TNG50 and TNG100 in Fig.~\ref{Compare}. The $\sigma_{\rm e}$ of TNG galaxies is calculated as the velocity dispersion of stellar particles along the X-axis within the 2D half mass radius described in \ref{density slope}. The line widths of the median value at high $\sigma_{\rm e}$ bins reflect the lack of massive galaxies in TNG due to the limited volume of the simulation box. It can be seen that the $\overline{\gamma}_{_{\rm T}}$ of TNG50 galaxies with $M_{200}>10^{13}\ {\rm M_\odot}$ and $\sigma_{\rm e}>150$ km/s are generally in good agreement with MaNGA, while results from TNG100 show obvious differences with them: for TNG100 galaxies with $M_{200}>10^{13}\ {\rm M_\odot}$ ($\sigma_{\rm e}>150$ km/s), their median $\overline{\gamma}_{_{\rm T}}$ decreases gradually from 2.03 (2.22) to 1.76 (1.76). However, this decreasing trend is consistent with \citet{Remus2017}, where ETGs selected from the Magneticum Pathfinder simulations \citep{Magneticum} were studied. At low mass end (at $\sigma_{\rm e} \sim$ 100 km/), both TNG50 and TNG100 overestimate the total density slope, by $\sim$0.4 for TNG50 and $\sim$0.37 for TNG100. To understand why do the TNG100 and TNG50 differ at high mass end, we further plot the $R_{\rm e}-M_{200}$, $R_{\rm e}-\sigma_{\rm e}$ and $R_{\rm e}-M_*$ relations in Fig.~\ref{50vs100}. The range of the horizontal axis for the comparison is constrained to be approximately the same as that in Fig.~\ref{Compare}. The $\lg R_{\rm e}$ of TNG100 is obviously higher by at least $0.2$ dex than TNG50, which results the observed lower value of $\overline{\gamma}_{_{\rm T}}$ for TNG100 in high-mass end in Fig.~\ref{Compare}.

In the related study \citetalias{Wang2023}, we measures the total density slope for a subset of central galaxies in groups and clusters by combining their stellar kinematics with weak gravitational lensing. For galaxies in the group bin with the best-fitting result at $\lg M_{200} [{\rm M_\odot}]$ of 13.2, $\overline{\gamma}_{_{\rm T}}$ of their mean density profile is $2.15^{+0.04}_{-0.05}$, which is in excellent agreement with the findings presented in this paper. However, for galaxies in the cluster bin with the best-fitting result at $\lg M_{200} [{\rm M_\odot}]$ of 13.92, $\overline{\gamma}_{_{\rm T}}$ of their mean density profile is $1.95^{+0.08}_{-0.09}$, which is lower than the result in this paper by about 0.2 for the same mass range. The source of this difference primarily stems from two aspects. First, the data points in Fig.~\ref{Compare} represent the stacked $\overline{\gamma}_{_{\rm T}}$, which is obtained by first averaging the density profiles of galaxies within a specific mass bin and then calculating the density slope as Eq.~\ref{eq: MWslope}. Through our tests we found that for pure JAM modelling results of MaNGA, the stacked $\overline{\gamma}_{_{\rm T}}$ in the $\lg M_{200} [{\rm M_\odot}]$ range of $13.5-14.8$ is lower than the median value of galaxy density slopes by 0.1. In Sec~\ref{sec:Stacked gamma}, we will further observe differences between calculating $\overline{\gamma}_{_{\rm T}}$ from averaged density profiles and obtaining median (or mean) $\overline{\gamma}_{_{\rm T}}$ values from the galaxies in the same bin. Second, the residual portion of the discrepancy can be attributed to differences in sample selection. \citetalias{Wang2023} include galaxies with Qual$\geq$0, whereas the results presented in this paper is based on galaxies with Qual$\geq1$. We have exclusively selected galaxies with ${\rm Qual}\geq1$ to ensure the reliability of our density slope measurements. \citetalias{Wang2023} employed mass measurements within a fixed radius (approximately 10-20 kpc) derived from \citetalias{Zhu2023a}, in conjunction with gravitational lensing and this calculation of total mass for MaNGA galaxies is relatively reliable even for galaxies with ${\rm Qual}=0$. The discrepancy in the measurement of $\overline{\gamma}_{_{\rm T}}$ between this paper and \citetalias{Wang2023} yields an intriguing implication: the total density profiles of galaxies are contingent upon their dynamical state. Galaxies with ${\rm Qual}\geq0$ exhibit relatively complex dynamical states and flatter density profiles. Furthermore, in Sec.~\ref{sec:Qual Distributin} we present the sample distribution characteristics of Qual on $\overline{\gamma}_{_{\rm T}}$ values to illustrate the systematic differences in the matter distribution among galaxies of different JAM modelling quality. If we calculate the stacked $\overline{\gamma}_{_{\rm T}}$ for the same group- and cluster-central galaxies, including those ${\rm Qual}=0$ sample, but solely based on the JAM modelling results, we also predict the consistent $\overline{\gamma}_{_{\rm T}}$, as Figure 9 of \citetalias{Wang2023} shows.

We also compare our observation with recent measurements of total density slope. Following the images captured by the Hubble Space Telescope, \citet{Etherington2023} employed the pixel-based strong lensing modelling technique, as detailed in \citet{PyAutoLens}, to derive lensing-only measurements of the logarithmic total density slope $\gamma$ for 42 early-type galaxies (ETGs) within the SLACS sample, which are originally presented in \citet{Bolton2008}. By jointly combining the weak and strong lensing constraints from 22 SLACS galaxies, \citet{Gavazzi2007} measured the virial-to-stellar mass ratio $M_{119}/M_* = 54^{+28}_{-21}$. $M*$ for 36 galaxies, assuming a Chabrier IMF, is available in \citet{Auger2010}, and we use these values to estimate the halo mass, $M_{200}$, for the SLACS sample. We employ a mass-concentration relation from \citet{Mh2c200} to convert $M_{119}$ to $M_{200}$. The $\gamma$ values obtained through pure lensing measurements exhibit a slight shallower trend when compared to the median result from JAM, but these measurements are generally consistent with the JAM results. 

For the high-mass end, we compare our result with ten massive galaxies from \citet{Newman2015} at the center of massive groups with an average mass of $M_{200} \sim 10^{14} {\rm M_\odot}$, and seven very massive galaxies from \citet{Newman2013a, Newman2013b} at the center of clusters have halo mass $M_{200} = 0.4-2 \times 10^{15} {\rm M_\odot}$, with total mass density profiles obtained via a combination of weak and strong lensing, resolved stellar kinematics and X-ray kinematics. The mass-weighted total density slopes $\overline{\gamma}_{_{\rm T}}$ of these massive galaxies reach $\sim 1.7$ for galaxy groups and $\sim 1.2$ for clusters. Comparing these results to our study, we find that the $\overline{\gamma}_{_{\rm T}}$ values for the ten group-scale central galaxies are notably lower than the mean value of MaNGA. However, they are in good agreement with the results from TNG100. 
Nevertheless, through semi-empirical approach, \citet{Shankar2017} found that some of their model predictions can match the variation of $\overline{\gamma}_{_{\rm T}}$ from SLACS galaxies to group-scale central galaxies in \citet{Newman2015}, but the density slopes of cluster-scale galaxies in \citet{Newman2013a} are still too low to be explained. They also demonstrated that the dependence of $\overline{\gamma}_{_{\rm T}}$ on halo mass is a genuine effect, only partially influenced by the increase in effective radius with stellar mass, reflecting the structural non-homology, where the relative density distribution of stars and dark matter vary systematically from isolated galaxies to central galaxies in clusters. The dependence of the total density slope on the environment cannot be ignored. Further investigations are needed to understand the origin of the discrepancy and to improve the accuracy of the total density slope measurements for galaxy groups and clusters.

\subsection{Dark matter and stellar components}
\label{Sec4.2}
\begin{figure}
    \centering
    \includegraphics[width=1\columnwidth]{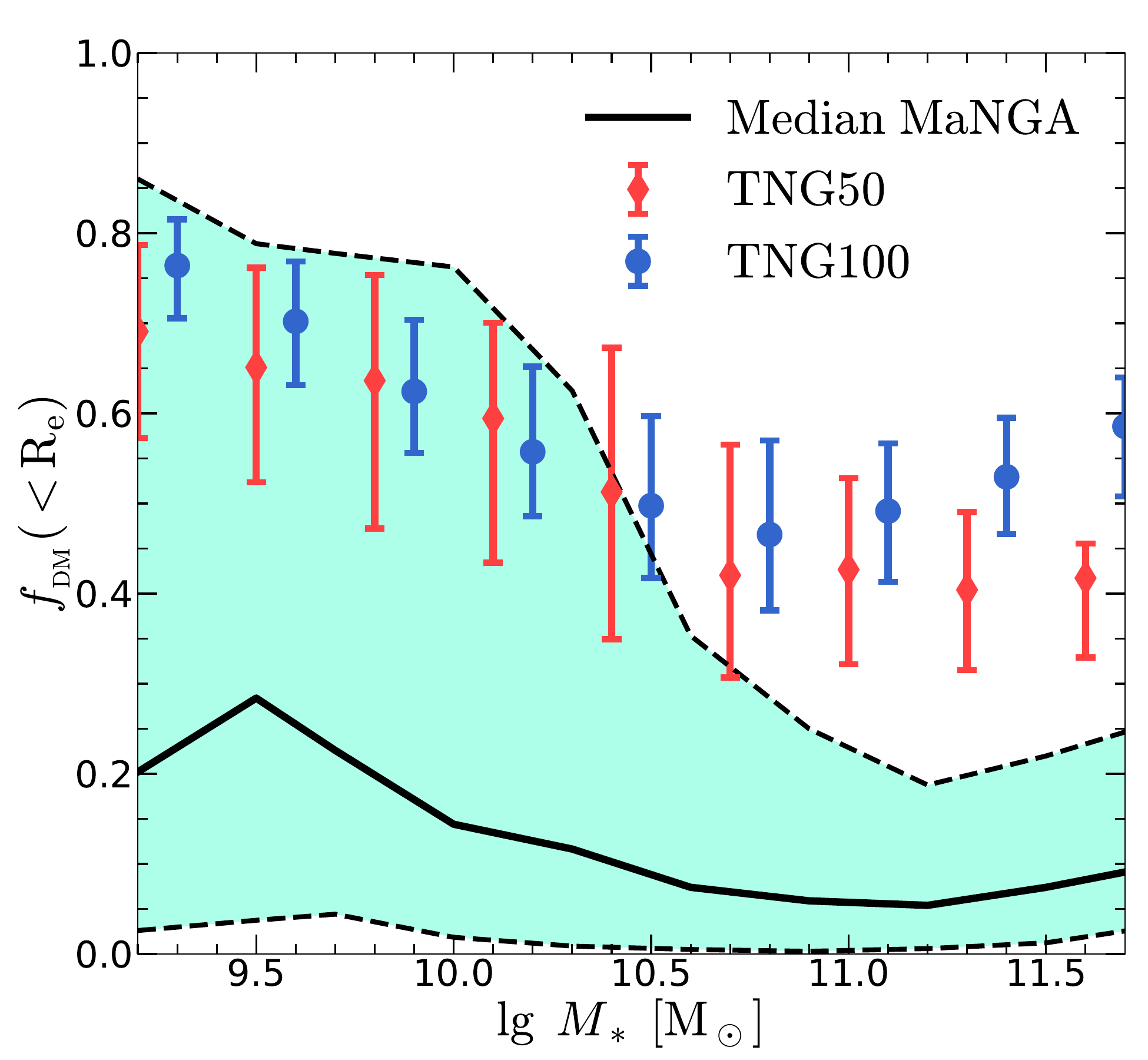}
    \caption{ $f_{\rm DM}(<R_{\rm e})$ as a function of $M_*$ for central galaxies. For MaNGA galaxies, the black solid line shows their median and the green shadow region indicates the 1-$\sigma$ scatter. For TNG50 and TNG100, the medians and 1-$\sigma$ scatters are showed by scattered dots with error bars.}
    \label{fdm_Ms}
\end{figure}

To inspect the reason of a relatively shallow $\overline{\gamma}_{_{\rm T}}$ in galaxies of IllustrisTNG simulation, we investigate the decomposed dark matter and stellar components respectively. The central dark matter fraction $f_{\rm DM}(<R_{\rm e})$ is defined as the mass ratio of dark matter over the total mass enclosed within the 2D effective radius $R_{\rm e}$ for both MaNGA and TNG galaxies. In \citetalias{Zhu2023b} (Figure 11) we already have looked into the $f_{\rm DM}(<R_{\rm e})$-$M_*$ relation. Here we show again the relation in Fig.~\ref{fdm_Ms} to compare it with the results from TNG. For MaNGA galaxies, according to the suggestion in Table 2 of \citetalias{Zhu2023a}, we add the filtering condition $|f_{\rm DM,cyl}-f_{\rm DM,sph}|<0.1$, to improve the accuracy of measuring the dark matter fraction. Although the trend of the $f_{\rm DM}(<R_{\rm e})$-$M_*$ relation is roughly same, the difference between MaNGA and TNG is obvious: simulated galaxies tend to have much more dark matter within their central region and this difference is greater at the large $M_*$ end, especially for TNG100. 

\begin{figure}
    \centering
    \includegraphics[width=1\columnwidth]{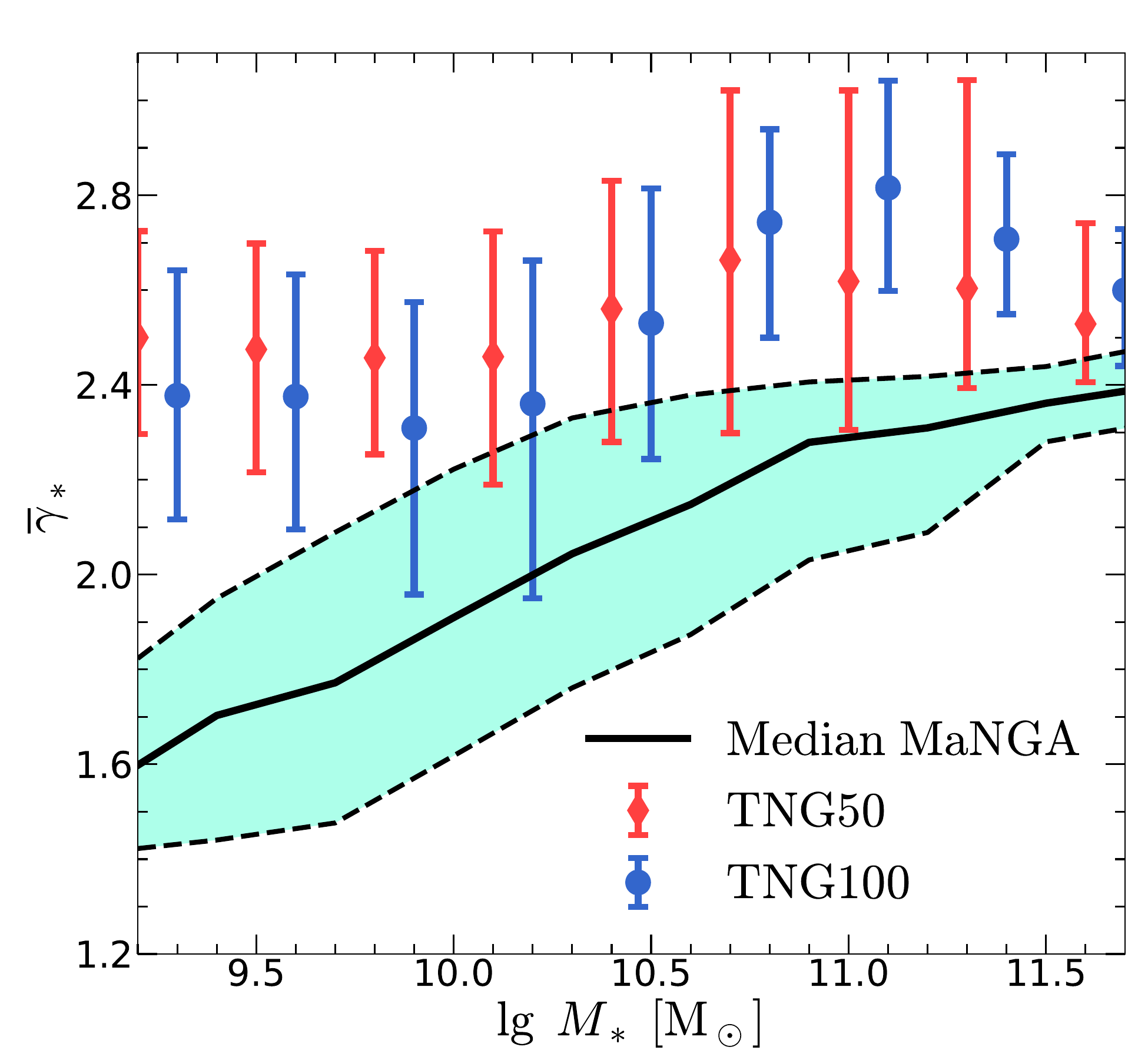}
    \caption{ $\overline{\gamma}_*$ as a function of $M_*$ for central galaxies. For MaNGA galaxies, the black solid line shows their median and the green shadow region indicates the 1-$\sigma$ scatter. For TNG50 and TNG100, the medians and 1-$\sigma$ scatters are showed by scattered dots with error bars.}
    \label{Gs_Ms}
\end{figure}

In Fig.~\ref{Gs_Ms} we compare the $\overline{\gamma}_*$-$M_*$ relation between MaNGA and TNG galaxies. It is evident that the numerical simulations predict much higher stellar density slopes compared to the observations. Therefore, the shallower overall density profiles of the simulated galaxies, as depicted in Fig. \ref{Compare}, can be attributed entirely to the disparity in the dark matter fraction between the simulations and observations, rather than differences in the stellar density distribution. Fig.~\ref{fdm_Ms} and Fig.~\ref{Gs_Ms} also demonstrate that although TNG50 appears to better predict the overall density profiles at the high-mass end, this is merely an incidental effect of the slightly flatter stellar density profiles and relatively lower dark matter fraction combined. It is not a robust indication of TNG50 outperforming TNG100 in modeling the total density profiles.

Previous lensing and dynamical observations find that massive elliptical galaxies have a mass density profile close to isothermal ($\rho\propto r^{-2}$, $\overline{\gamma}_{_{\rm T}}=2$) within an effective radius \citep[][]{Koopmans2009, Auger2010, Tortora2014, Poci2017, LiRui2018, LiRan2019} or well beyond the effective radius \citep{Gavazzi2007}. This phenomenon is commonly referred to as the ‘bulge-halo conspiracy’ \citep[e.g.,][]{Dutton2014}. Given that neither baryons nor dark matter exhibit an isothermal density profile on their own, it suggests that there is an interaction or ‘conspiracy’ between baryons and dark matter that collectively results in the formation of an isothermal density profile.

\begin{figure*}
    \centering
    \includegraphics[width=2\columnwidth]{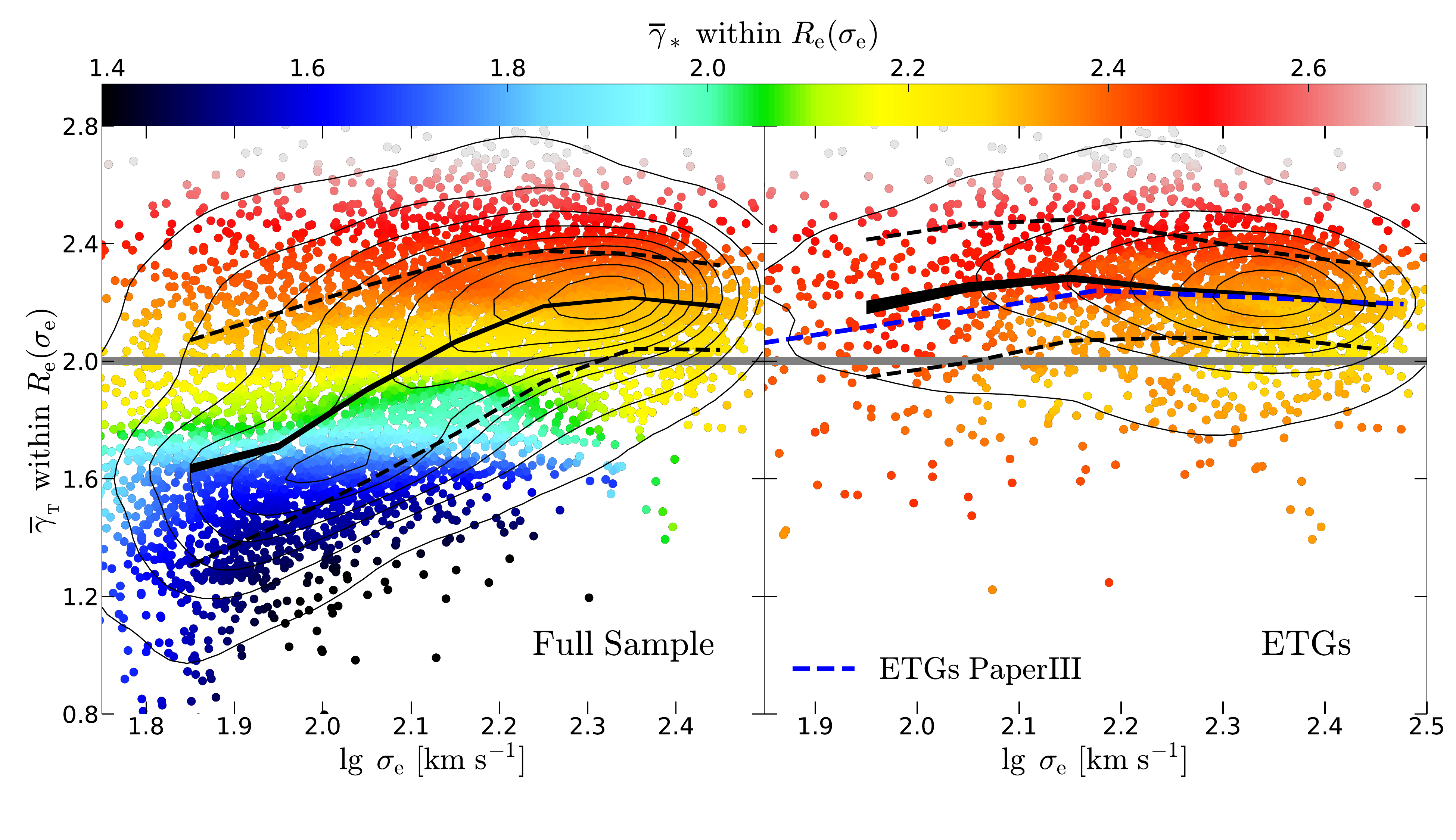}
    \caption{ $\overline{\gamma}_{\rm_{\rm T}}$ within $R_{\rm e}(M_*)$ as a function of $M_*$, colored by \textsc{loess}-smoothed $\overline{\gamma}_*$ within $R_{\rm e}(M_*)$ (\texttt{frac} = 0.05), for full sample (left) and for ETGs. In each panel, the black line shows the median with the line width represents the stand error of the median, and the region enclosed by dashed lines indicates the 1-$\sigma$ scatter. The blue dashed line marks the best fit for ETGs in \citetalias{Zhu2023b}. The black contours in each panel show the kernel density estimate for the sample distribution. The slope value of the isothermal profile is shown with a grey horizontal line.}
    \label{conspiracy1}
\end{figure*}

\begin{figure*}
        \centering
        \includegraphics[width=2\columnwidth]{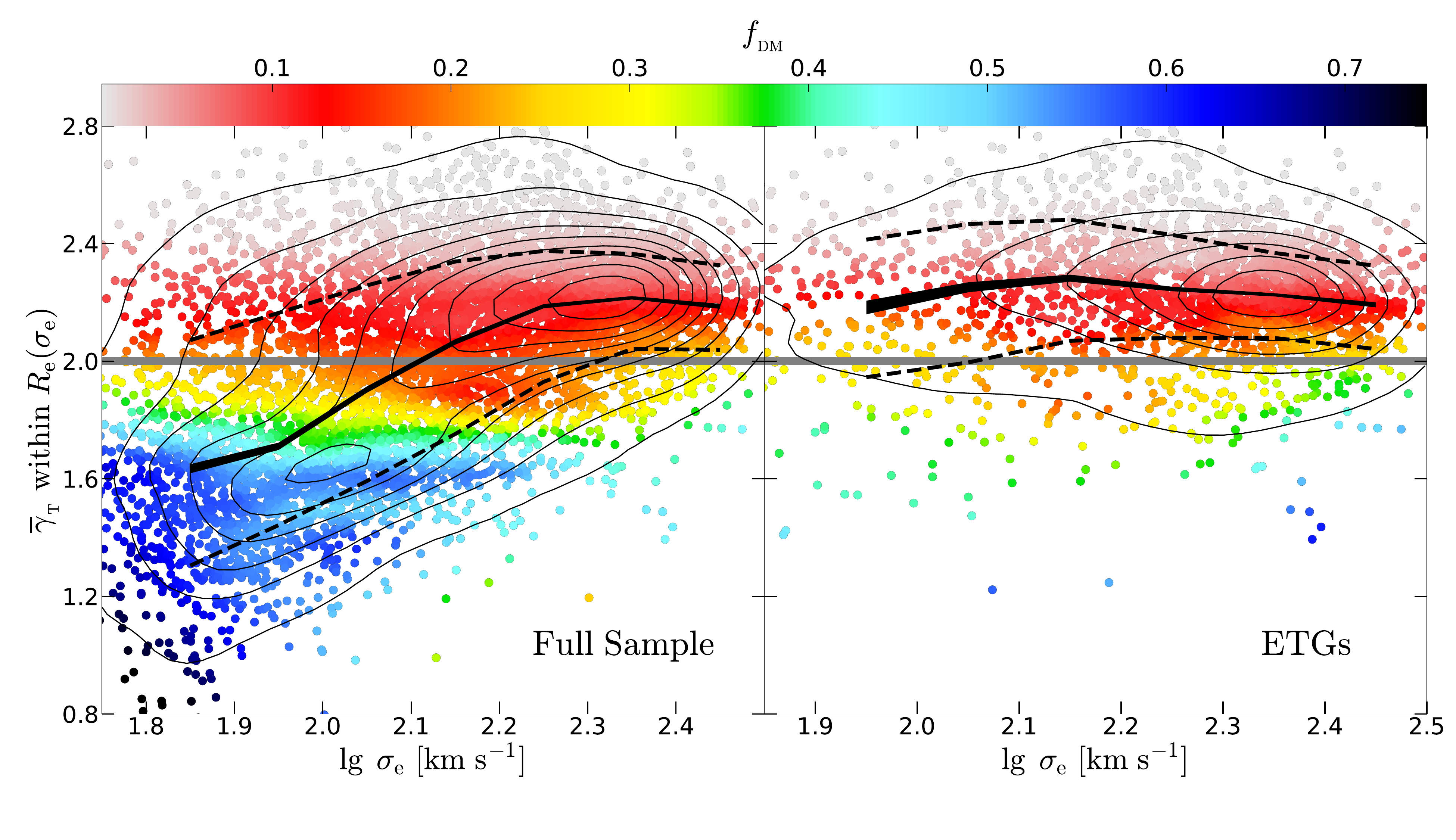}
        \caption{ Same as Fig.~\ref{conspiracy1}, but colored by \textsc{loess}-smoothed $f_{\rm DM}(<R_{\rm e})$ (\texttt{frac} = 0.05).}
        \label{conspiracy2}
    \end{figure*}

We intend to use our results from MaNGA to elucidate this issue. Specifically, we examine the characteristics of the distribution of stars and dark matter in galaxies with respect to $\overline{\gamma}_{_{\rm T}}$ and $\sigma_{\rm e}$ in Fig.~\ref{conspiracy1} and Fig.~\ref{conspiracy2}, with particular focus on galaxies that exhibit a higher velocity dispersion while maintaining a total density slope of $\sim$2, to investigate how their stellar and dark matter distributions vary. This will allow us to explore whether there is a matching mechanism between luminous and dark component that contributes to the ‘bulge-halo conspiracy’. We utilize $\overline{\gamma}_*$ and $f_{\rm DM}(<R_{\rm e})$ to characterize the distribution of stellar and dark matter, respectively. 
For galaxies within a specified $\sigma_{\rm e}$ value, a smaller effective radius corresponds a steeper density profile within $R_{\rm e}$, leading to a higher $\overline{\gamma}$. To mitigate the impact of the diversity in $R_{\rm e}$ among galaxies and ensure a fair comparison, we employ the median $R_{\rm e}$-$\sigma_{\rm e}$ relation to derive an interpolation function, $R_{\rm e}(\sigma_{\rm e})$. Then we recompute $\overline{\gamma}$ and label them as ‘$\overline{\gamma}$ within $R_{\rm e}(\sigma_{\rm e})$’. For brevity, we will continue to refer to them as $\overline{\gamma}_{_{\rm T}}$ and $\overline{\gamma}_*$ in the following text.

To capture the overall trend of the data distribution, the colored scattering points have been smoothed using the \textsc{loess} technique, a Locally Weighted Regression method developed by \citet{Cleveland1979}, and implemented in the \textsc{loess} Python procedure\footnote{Available from \url{https://pypi.org/project/loess/}} created by \citet{Cappellari2013}. For full sample, the dispersion of $\overline{\gamma}_{_{\rm T}}$ is not random but instead depends on the stellar density profile and the proportion of dark matter. Given the stellar mass, a higher density slope of the stellar component corresponds to a higher total density slope of the galaxy. Similarly, a higher proportion of dark matter leads to a flatter total density profile. There is no sign that stellar and dark matter components balance each other to yield a $\overline{\gamma}_{_{\rm T}}\sim2$, as predicted by the bulge-halo conspiracy theory. But ETGs exhibit a markedly different scenario: across the entire $\sigma_{\rm e}$ range, the median value of $\overline{\gamma}_{_{\rm T}}\sim2$ remains constant at 2.2. The trend of change is even flatter than the fitting result of $\overline{\gamma}_{_{\rm T}}$-$\sigma_{\rm e}$ relations plotted in Fig.~\ref{scaling} because here we have corrected for the varying impact of $R_{\rm e}$ on the slope at the same $\sigma_{\rm e}$ and consider it a fairer comparison for $\overline{\gamma}_{_{\rm T}}$. This appears to corroborate the conspiracy theory regarding massive elliptical galaxies, if we disregard that $\overline{\gamma}_{_{\rm T}}$ is around 2.2 (rather than 2) and that their $\overline{\gamma}_*$ and $f_{\rm DM}(<R_{\rm e})$ still exhibit a systematic layered structure. 

\subsection{Stacked density profile}
\label{sec:Stacked gamma}
\begin{figure*}
    \centering
    \includegraphics[width=2\columnwidth]{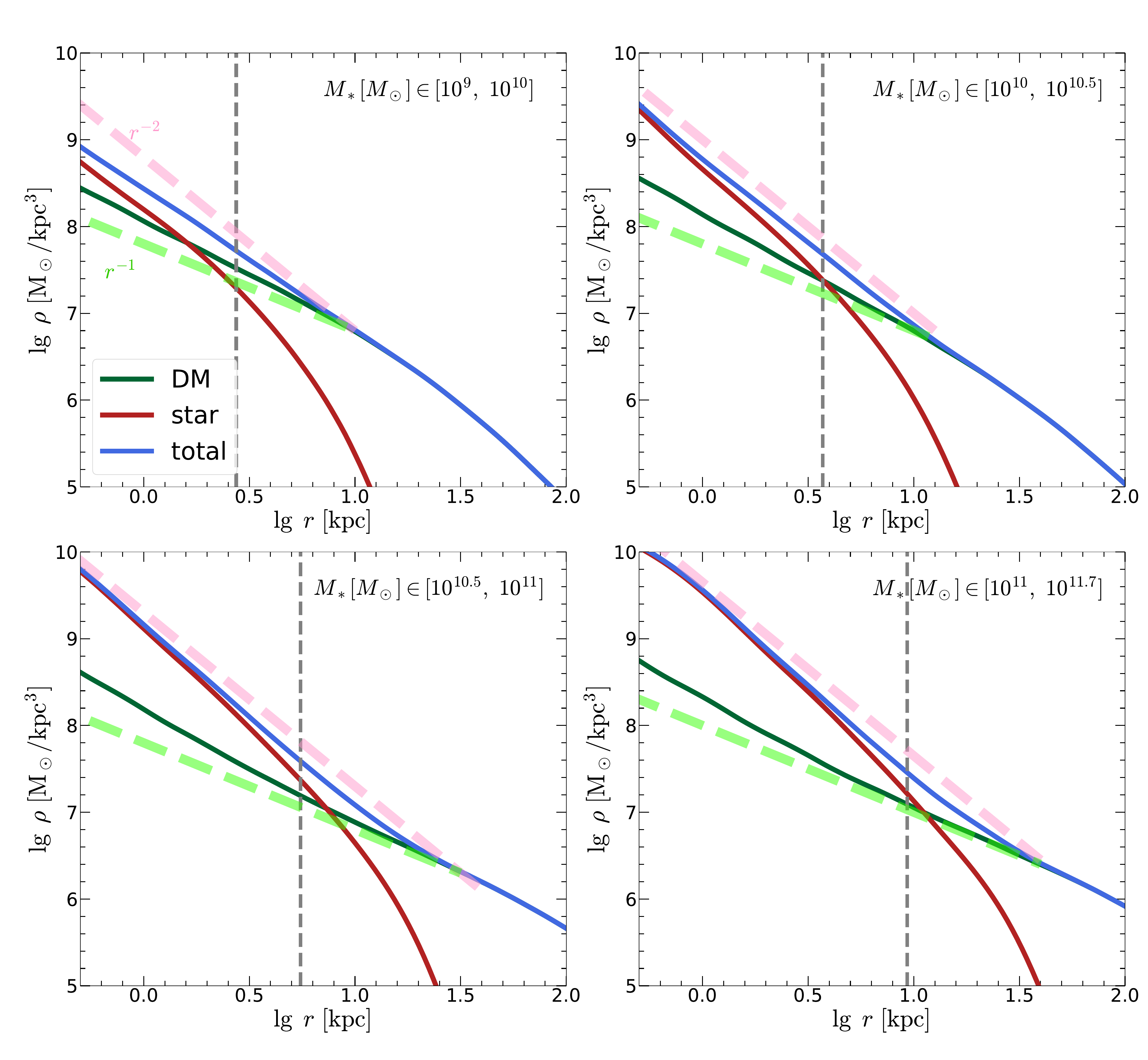}
    \caption{The stacked profile of MaNGA central galaxies, divided by $M_*$ as marked in each panel. The light green and pink dashed lines indicate the profile with 1 and 2 slope, respectively. The grey vertical dashed line marks the average of $R_{\rm e}$ in each mass bin.}
    \label{stkPro_Ms}
\end{figure*}

\begin{figure*}
    \centering
    \includegraphics[width=2\columnwidth]{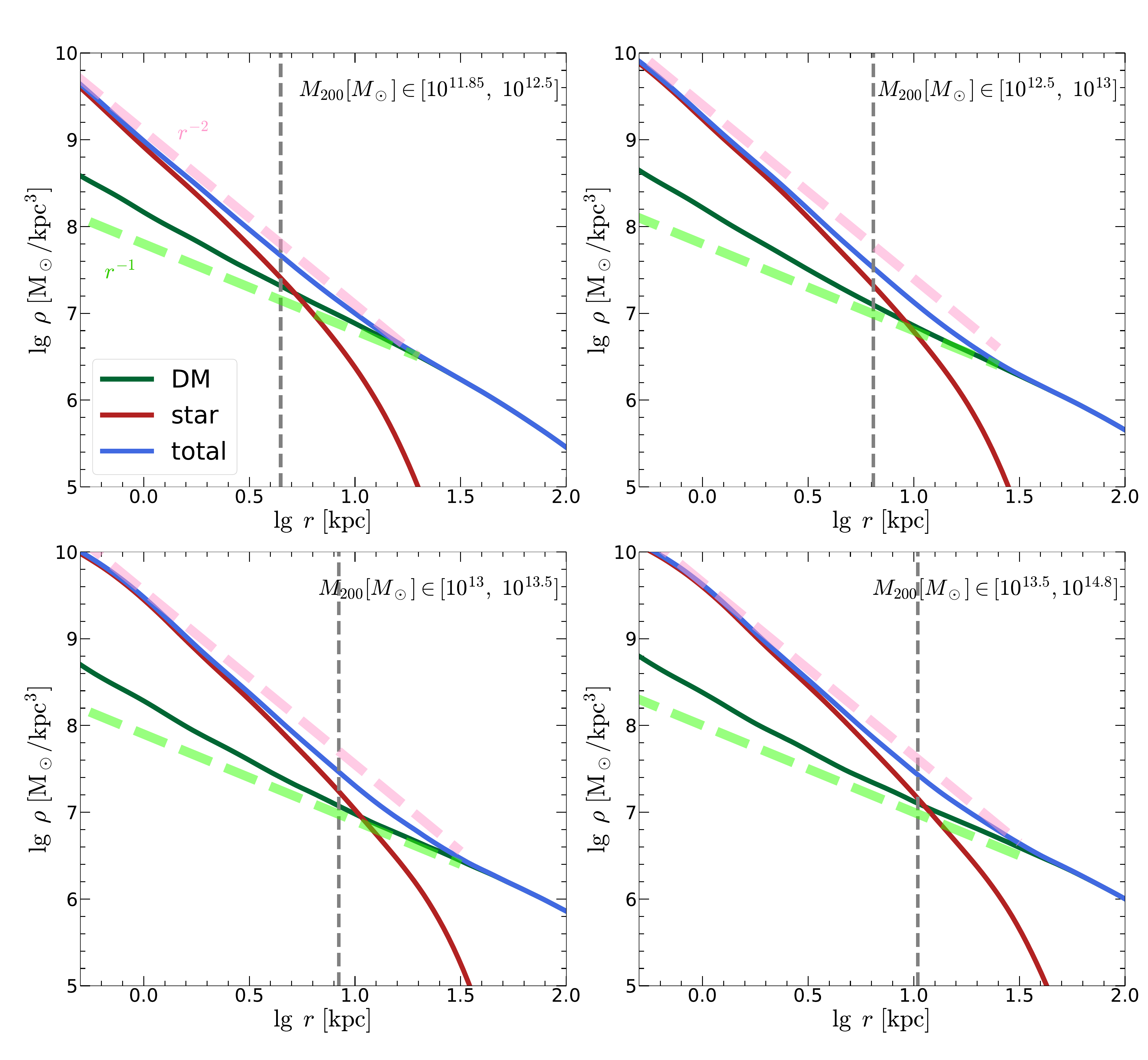}
    \caption{Same as Fig.~\ref{stkPro_Ms}, but the mass bin is split by $M_{200}$.}
    \label{stkPro_Mh}
\end{figure*}

\begin{figure*}
    \centering
    \includegraphics[width=2\columnwidth]{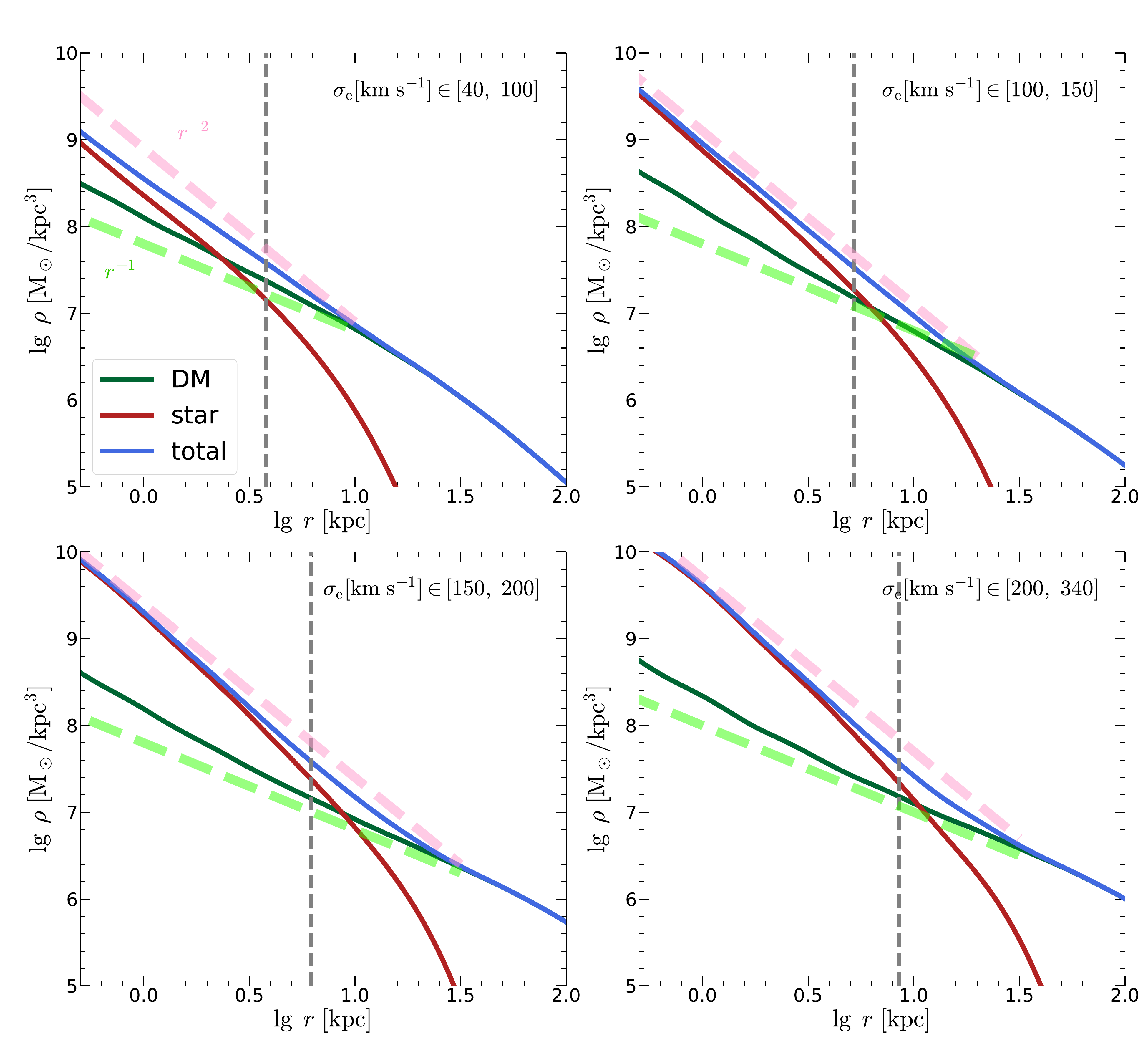}
    \caption{ Same as Fig.~\ref{stkPro_sigma}, but the mass bin is split by $\sigma_{\rm e}$.}
    \label{stkPro_sigma}
\end{figure*}

Next, we stack (take the average of) the density profiles of the central galaxies in different mass bins, and calculate the slopes of the stacked density profiles of stars, dark matter, and total matter to study how the stacked density slope varies with mass. Firstly, we divide the MaNGA galaxies into sub-sample bins, based on $M_*$, $\sigma_{\rm e}$ and $M_{200}$, respectively. Then we obtain the mean density profiles of stars, dark matter, and total matter for each mass bin, which are plotted in Fig.~\ref{stkPro_Ms}, Fig.~\ref{stkPro_Mh} and Fig.~\ref{stkPro_sigma}. In these plots, each panel represents one mass bin, and the dashed gray vertical line indicates the mean $R_{\rm e}$ of galaxies in that bin. Note that, \citetalias{Yang2007} only measured the reasonable dark matter halo mass of galaxies with $M_*$ bigger than $\sim 10^{10}\ {\rm M_\odot}$ and the density profiles of galaxies in this mass bin corresponding to the upper left panel of Fig.~\ref{stkPro_Ms} are almost absent in Fig.~\ref{stkPro_Mh}. Once again, we observe that the stellar matter dominates within $R_{\rm e}$, especially in galaxies with larger masses, where the dark matter density in the central region is only 1/10 of the stellar density. We can see that within the mean $R_{\rm e}$, the slope of the stacked dark matter density profile in each mass bin is slightly steeper than the prediction of NFW profile, namely 1, while the slope of the stacked total matter density profile, except for the lowest $M_*$ bin, is very close to 2 in all other bins. 

\begin{figure*}
    \centering
    \includegraphics[width=2\columnwidth]{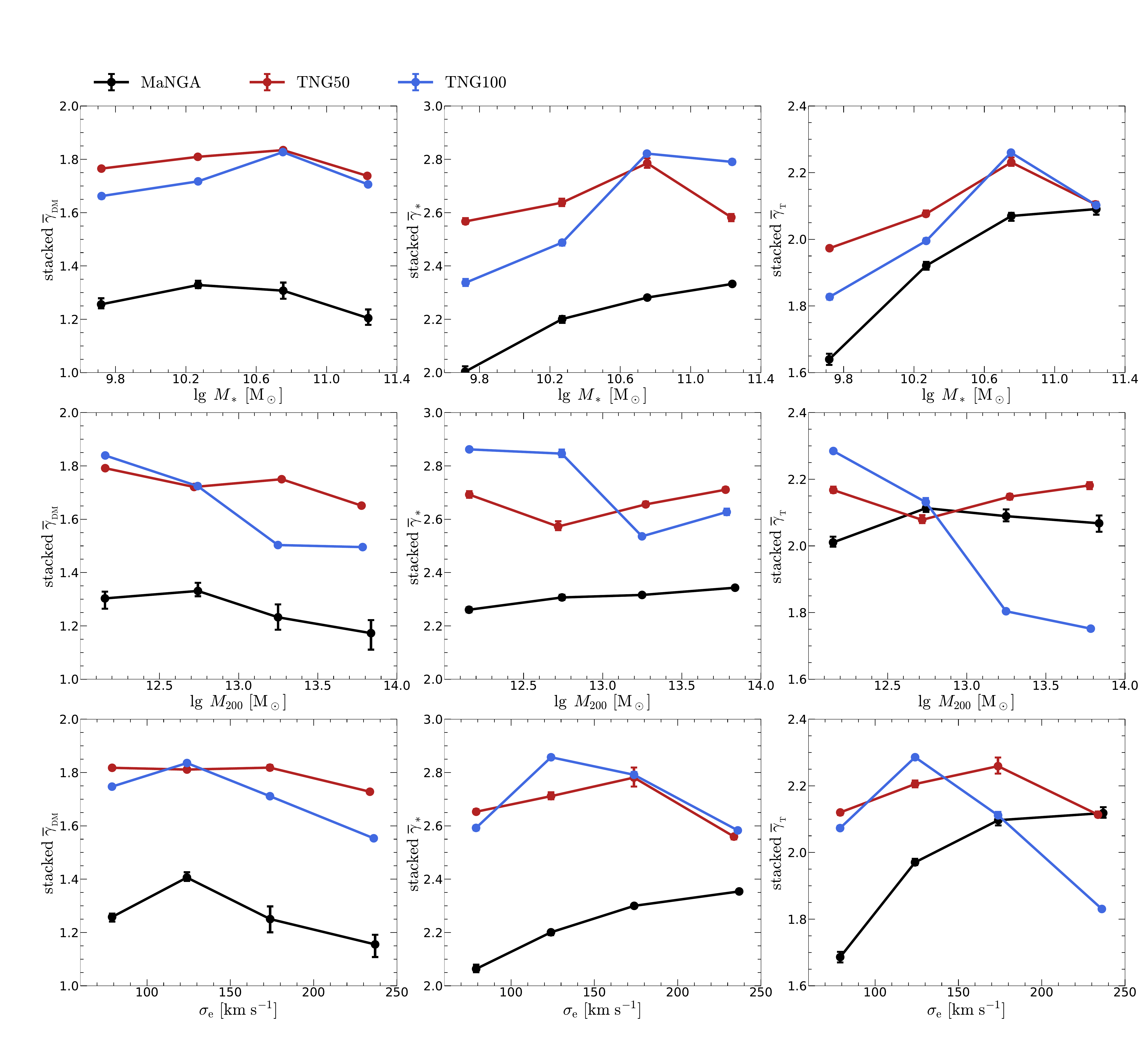}
    \caption{ \textbf{Top}: stacked $\overline{\gamma}_{_{\rm DM}}$, $\overline{\gamma}_*$ and $\overline{\gamma}_{_{\rm T}}$ (from left to right) as a function of the mean $M_*$ for each stellar mass bin. The results of MaNGA, TNG50 and TNG100 are plotted by filled circles with error bars indicating the 1-$\sigma$ scatter. \textbf{Middle}: the same as the top panels, but plotted as functions of the mean $M_{\rm 200}$ for each halo mass bin. \textbf{Bottom}: the same as the top panels, but plotted as functions of the mean $\sigma_{\rm e}$ for each velocity dispersion bin.}
    \label{stkSlope}
\end{figure*}

We investigate how the stacked density slope changes with galaxy properties. In Fig.~\ref{stkSlope}, we show how the stacked density slope of different components in MaNGA and TNG galaxies varies as a function of the mean mass (velocity dispersion) of galaxies within each stacked mass (velodicty dispersion) bin. We used bootstrap re-sampling to obtain the uncertainty of stacked density slopes within each bin. Specifically, we perform a re-sampling with replacement of the galaxies in each bin. For each re-sampling, we select galaxies from the bin with replacement and stack their density and enclosed mass profiles to calculate the stacked density slope. The number of galaxies sampled in each re-sampling is equal to the total number of galaxies contained in that bin. We repeat this process 100 times to obtain the standard deviation of the stacked density slopes. 

For a fair comparison, we also resample galaxies from TNG50 and TNG100 to match the $M_*$, $M_{\rm 200}$ and$\sigma_{\rm e}$ space distribution of galaxies of MaNGA galaxies in each subsample bin. After obtaining the re-sampled TNG galaxies, we computed the stacked density slopes. However, it should be noted that due to the limited volume of the simulation box, the massive galaxy subsample in TNG50 and TNG100 may be less representative. For the mass range of $M_{200}>{10}^{13}\ {\rm M_\odot}$, where TNG50 (TNG100) only has 24 (127) galaxies, while MaNGA has 1081 galaxies. 

In Fig.~\ref{stkSlope}, the stacked density slopes of different components are denoted as stacked $\overline{\gamma}_{_{\rm DM}}$, stacked $\overline{\gamma}_*$, and stacked $\overline{\gamma}_{_{\rm T}}$, respectively. For MaNGA galaxies, overall, stacked $\overline{\gamma}_{_{\rm T}}$ increases with increasing $\sigma_{\rm e}$ and $M_*$ and become flatter in the highest two $\sigma_{\rm e}$ and $M_*$ bins. The behavior of stacked $\overline{\gamma}_{_{\rm T}}$ as a function of $M_{200}$ is relatively flat, which is due to the lower limit cutoff of our dark halo mass at $10^{12}\ {\rm M_{\odot}}$, not including central galaxies with smaller $M_*$ or $\sigma_{\rm e}$.

Different with the situation of median value in Fig.~\ref{scaling} that $\overline{\gamma}_*$ is uniformly higher by $\sim0.2$ than $\overline{\gamma}_{_{\rm T}}$ across nearly all MaNGA samples, stacked $\overline{\gamma}_*$ is higher than stacked $\overline{\gamma}_{_{\rm T}}$ by 0.36, 0.28, 0.21 and 0.24 in the four $M_*$ bins (by 0.38, 0.23, 0.20 and 0.24 in the four $\sigma_{\rm e}$ bins), respectively. We find that this is because at the low-$M_*$ ($\sigma_{\rm e}$) end, the value of stacked $\overline{\gamma}_*$ (slope of the mean stellar profile) is higher by $\sim0.2$ than the median of $\overline{\gamma}_*$ while at the high-$M_*$ (high-$\sigma_{\rm e}$) end these two values match quite well, and the difference of these two values decreases with the increasing $M_*$ or $\sigma_{\rm e}$. This results in a less pronounced growth trend of stacked $\overline{\gamma}_*$ with $M_*$ or $\sigma_{\rm e}$ compared to stacked $\overline{\gamma}_{_{\rm T}}$. We attempt to explain the reasons for the different behaviors of stacked $\overline{\gamma}_*$ in different $M_*$ or $\sigma_{\rm e}$ bins in Appendix \ref{Appx.A}. The dark matter density slope is slightly steeper than $\gamma\sim1$, ranging between 1.1-1.4, presenting a flat inverted-U shape with changes in $M_*$ and $\sigma_{\rm e}$, peaking at $\sim1.8\times10^{10}M_{\odot}$ ($\sigma_{\rm e}\approx150$ km/s), and tending towards $\gamma\sim1$ at both the higher and lower mass ends.

There are significant differences in the values of  stacked $\overline{\gamma}_{_{\rm DM}}$ and stacked $\overline{\gamma}_*$ between MaNGA and TNG galaxies, with TNG galaxies having higher values than MaNGA galaxies. Although both the decomposed stellar and dark matter components in the simulations show obvious higher density slopes, the relationship between total density slopes and stellar mass in TNG50 and TNG100 appears to produce a similar trend and slightly higher magnitude to observations. This is because the dark matter fraction in galaxies in the TNG simulations is evidently higher than those observed in MaNGA. The relatively flat dark matter density distribution substantially counterbalances the steep stellar density distribution, coincidentally resulting in a similar outcome to the observations.

%\textcolor{red}{XXXX May need discussion.XXX As $M_*$ increases, the difference in stacked $\overline{\gamma}_{_{\rm T}}$ between MaNGA and TNG galaxies decreases, which reflects that the fraction of dark matter in TNG galaxies is significantly higher than that in MaNGA galaxies at the high $M_*$ end, which effectively reduces stacked $\overline{\gamma}_{_{\rm T}}$ and achieves results that are more consistent with MaNGA galaxies. The stacked $\overline{\gamma}_{_{\rm DM}}$ of MaNGA galaxies is generally higher than 1, but gradually decreases with increasing $M_*$ and $M_{200}$ in the high mass regime, while stacked $\overline{\gamma}_*$ steadily increases with increasing $M_*$ and $M_{200}$, indicating that the distribution of dark matter is influenced by the stellar component, and its density slope is compressed to become steeper by the gravitational field of the stellar component. As the mass of the galaxy increases, the increased baryonic matter produces stronger feedback from the central black hole and supernovae, resulting in the outward expansion of the dark matter and reducing the slope of the dark matter density profile in the central region of the galaxy.}

For the relationship between $\sigma_{\rm e}$ and total density slope, and for the relationship between $M_{\rm 200}$ and total density slope, the results given by TNG50 and TNG100 are quite different, and they are significantly different from the observations. TNG100 predicts significantly lower total density slopes at the high-velocity dispersion end compared to TNG50, with a difference reaching 0.2-0.4. In fact, we find that for the central galaxies in large mass halos in TNG100, they have flatter stacked stellar and dark matter density profiles, and also have a larger dark matter fraction (Fig.~\ref{fdm_Ms}). This is why it gives a flatter density profile.

It is worth noting that the TNG simulations and MaNGA observations also have different $\sigma_{\rm e}$-$M_*$ relationships. At the same stellar mass, MaNGA galaxies have a larger velocity dispersion. This is also a reason why the results are complex in the comparisons between TNG simulations and MaNGA in different density slope scaling relationships.

\section{Discussion}
\label{discussion}
\subsection{Bulge-halo conspiracy}
In Fig.~\ref{scaling} we illustrate that the median $\overline{\gamma}_{_{\rm T}}$ of ETGs remains at 2.2 only with $\sigma_{\rm e}\ga150\ \kms$. This is the same value originally reported as ‘universal’ for ETGs by \citet{Cappellari2015}. Below this velocity dispersion, $\overline{\gamma}_{_{\rm T}}$ becomes more dispersed and decreases as $\sigma_{\rm e}$ decreases. This finding corroborates the conclusions from the dynamical studies of ETGs in previous ATLAS$^{\rm 3D}$ and early MaNGA data releases \citep{Poci2017, LiRan2019}. In Fig.~\ref{conspiracy1} and Fig.~\ref{conspiracy2}, we observe that when we keep $R_{\rm e}$ fixed to calculate the $\overline{\gamma}_{_{\rm T}}$ of ETGs for a specific $\sigma_{\rm e}$ value, the median $\overline{\gamma}_{_{\rm T}}$ remains consistently at 2.2, regardless of changes in $\sigma_{\rm e}$. This result seems to support the bulge-halo conspiracy theory. However, this situation does not persist if we employ different methods of ETGs classification. In the preceding sections, we have utilized the stringent selection criteria proposed by \citet{MDL} to identify our ETG sample, which predominantly consists of elliptical (E) and lenticular (S0) galaxies. The primary criteria for this morphological classification of ETGs are ${\rm P_{LTG}}<0.5$ and T-Type<0, effectively distinguishing ETGs from LTGs.

Following \citet{classify1} and \citet{classify2}, the luminosity-morphology parameter $L_{\rm dev}/L_{\rm tot}$ and the specific star formation rate (sSFR) is individually adopted by us. $L_{\rm dev}/L_{\rm tot}$ is defined as the ratio of the de Vaucouleurs luminosity to the total galaxy luminosity and characterizes the galaxy's fraction of the elliptical component, while sSFR is defined as the ratio of the star formation rate (in ${\rm M_\odot/Gyr}$) to the galaxy stellar mass. ETGs typically have the elliptical morphology ($L_{\rm dev}/L_{\rm tot}>0.5$) and quenched star formation ($\lg {\rm sSFR/Gyr^{-1}}\leqslant-2$), while late-type galaxies are dominated by disks ($L_{\rm dev}/L_{\rm tot}<0.5$) and still have ongoing star formation ($\lg {\rm sSFR/Gyr^{-1}}\geqslant-1.5$). $L_{\rm dev}/L_{\rm tot}$ comes from the MaNGA Deep Learning morphological catalogue, while sSFR is computed using the reliable (‘QCFLAG’=0) SFR estimation derived from $\rm H\alpha$ luminosity (‘log\_SFR\_$\rm H\alpha$’) and the photometric stellar mass (‘log\_Mass’) sourced from the pyPipe3D analysis catalogue, which is based on SDSS DR17 \citep{pyPipe3d}. 

\begin{figure*}
    \centering
    \includegraphics[width=2\columnwidth]{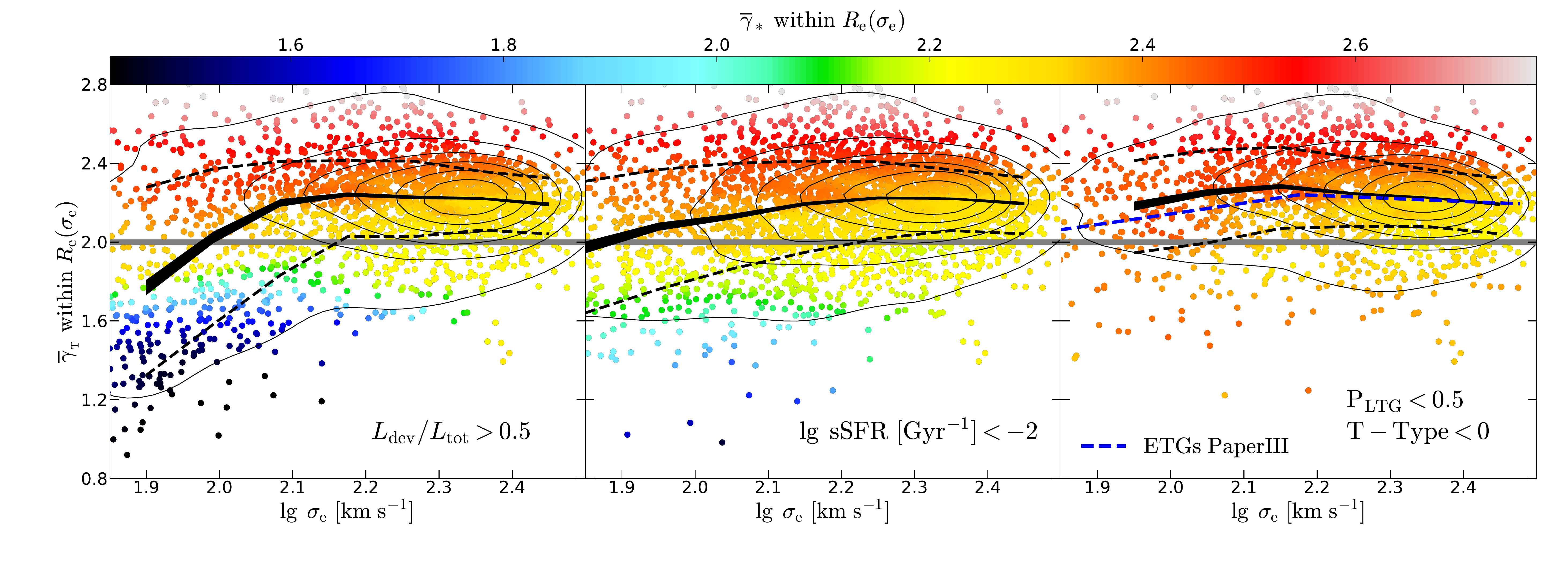}
    \caption{Same as Fig.~\ref{conspiracy1}, but we employ three distinct ETGs classification methods, indicated from left to right as the luminosity-morphology parameter, the specific star formation rate, and the morphological classification, each with its respective criterion listed in the corresponding panel.}
    \label{Compare_Classify}
\end{figure*}

We present the results for ETGs once again in Fig.~\ref{Compare_Classify}, this time comparing the differences brought by different classification methods. Under the criterion of luminosity-morphology parameter, sSFR and morphological classification method, there are about 2.2k, 3k and 2.2k ETGs, respectively. Clearly, apart from the differences in quantity, these three classification methods also exhibit distinct distributions in the corrected $\overline{\gamma}_{_{\rm T}}$-$\sigma_{\rm e}$ parameter space for ETGs. Under the luminosity-morphology parameter classification, ETGs display the most scattered density slopes, with a median $\overline{\gamma}_{_{\rm T}}$ value dropping to 1.7 at $\lg \sigma_{\rm e}=1.8$, nearly same as the changing trend of the full sample. The total density slope values of ETGs obtained under the sSFR classification exhibit a more compact distribution, with the median $\overline{\gamma}_{_{\rm T}}$ decreasing from 2.2 to 1.9 as $\sigma_{\rm e}$ decreases, but this trend is still more prominent compared to that observed under the morphological classification.

As mentioned in \citet{classify2}, the luminosity-morphology parameter classification is not accurate for S0 galaxies, and different classification methods have their respective limitations. However, here we aim to illustrate that the distribution of stars and dark matter in ETGs always exhibits clear layering with respect to the variation in $\overline{\gamma}_{_{\rm T}}$ and the difference lies in the range delineated by different classification methods in the galaxy $\overline{\gamma}_{_{\rm T}}-\sigma_{\rm e}$ parameter space.

Disregarding the influence of classification methods and solely focusing on the distribution of stars and dark matter within galaxies, for high-velocity-dispersion MaNGA galaxies, while the median $\overline{\gamma}_{_{\rm T}}$ at a given velocity dispersion approaches $\overline{\gamma}_{_{\rm T}}\sim 2$, our results for the separated stellar density profile and dark matter density profile reveal that these two components do not ‘conspire’ to form an isothermal total density profile. To the contrary, as the stellar density deviates more from the average level, the fraction of dark matter in the galactic center decreases, and the total density profile deviates further from the isothermal (Fig.~\ref{conspiracy1} and Fig.~\ref{conspiracy2}). The outcome of isothermal like density slope for early type galaxies appears to be a coincidental phenomenon in the mass growth process of galaxy-dark halo systems within a given velocity dispersion bin.

\subsection{Discrepancy with hydrodynamical simulation}
\label{Sec:IMF}
The scaling relationships for the total density profiles slopes of MaNGA galaxies obtained in this study cannot be replicated in the magneto-hydrodynamic cosmological TNG simulation. As shown in Fig.~\ref{Compare}, for galaxies at high $M_{\rm 200}$ and $\sigma_{\rm e}$ end, the TNG100 simulation underestimates the total density slope by $\sim 0.4$, while at low $\sigma_{\rm e}$ end, both TNG100 and TNG50 significantly overestimate the total density slope of the galaxies. The differences can be partially attributed to the different $M_{200}-R_{\rm e}$ and $\sigma_{\rm e}-R_{\rm e}$ relations between observations and TNG simulations (particularly evident in TNG100), but cannot be solely attributed to this factor. To correct the difference caused by $R_{\rm e}$, we derive interpolation functions for $R_{\rm e}(M_{200})$ and $R_{\rm e}(\sigma_{\rm e})$ based on the $R_{\rm e}-M_{200}$ and $R_{\rm e}-\sigma_{\rm e}$ relations of MaNGA galaxies and recalculate the density slope with the fixed $R_{\rm e}$ for TNG50 and TNG100 galaxies as discribed in Sec.~\ref{Sec4.2}. The results of TNG50 are basically unchanged because the variation of its $R_{\rm e}$ is similar to that of MaNGA. For TNG100, the difference on $\overline{\gamma}_{_{\rm T}}$ with MaNGA at the high-mass end reduce from larger than 0.4 to smaller than 0.2, while in the low-mass range the difference still exists and is significant. The simulated trends in the $\overline{\gamma}_{_{\rm T}}$-$M_*$ (although we do not show in the text) and the $\overline{\gamma}_{_{\rm T}}$-$M_{200}$ relationships also differ markedly from the observations. The discrepancy becomes even larger when we separate and study the stellar density and dark matter components. Both the stellar density slope and the dark matter density slope are higher in numerical simulations. But why do the total density slopes of observed MaNGA galaxies, particularly those of larger mass, exceed those in the TNG100 simulation, despite the fact that their decomposed components exhibit lower slopes? The key lies in the fact that the proportion of dark matter in the central regions in the numerical simulations is significantly higher than in the observations.(refer to Fig.~\ref{Compare}, Fig.~\ref{fdm_Ms} and Fig.~\ref{stkSlope}). The issue regarding the overestimation of the dark matter fraction in hydrodynamical simulations has been highlighted in previous studies. For the original Illustris simulation, \citet{Xu2017} found dark matter on average contributes about 40 – 50 per cent to the (projected) total matter distributions at the centres of ETGs, which is higher than suggested by the observational results. \citet{Lovell2018} found a clear tension between the high dark matter fraction in TNG100 central galaxies and the low dark matter fractions within the central regions of the ATLAS$^{\rm 3D}$ elliptical galaxies. In the present study, we confirm this discrepancy by utilizing a larger sample spanning a broader mass range.

Theoretically, AGN feedback could alter the slope of a galaxy's density profile. For instance, \citet{Peirani2019} discovered that incorporating AGN feedback in hydrodynamical simulations is essential for achieving improved agreement with observational values and the variation trends of $\overline{\gamma}_{_{\rm T}}$. They also found the derived slopes are slightly lower than in the observational values when AGN is included because the simulated galaxies tend to be too extended, especially the least massive ones. \citet{WangTNG1} demonstrated that the kinematic wind feedback from AGNs could flatten the density profiles of ETGs and weakening the AGN feedback in numerical simulations might bring the resulting total density profiles closer to the observations of massive galaxies. However, given that the stellar density profiles of the simulated galaxies are already steeper than the observations as we see in Fig.~\ref{Gs_Ms}, weakening the AGN feedback in the simulations would only exacerbate this difference. Therefore, it is hard to imagine that merely reducing AGN feedback could fully resolve the discrepancies we observed in this study.

Indeed, a more critical factor causing the discrepancy between simulations and observations might lie in the assumption of an invariant IMF in the star formation processes within the TNG simulations. Within TNG, regardless of galaxy properties, the IMF consistently adheres to a bottom-light form as described by the Chabrier formula. However, previous studies suggest that the IMF of star formation could vary depending on the characteristics of the galaxy. Specifically, in ETGs with high velocity dispersion, the IMF may skew closer to a bottom-heavy situation \citep[e.g.][]{Dokkum&Conroy2010, Cappellari2012}. Consequently, the stellar mass of high-velocity-dispersion galaxies in simulations might be underestimated, leading to a higher derived dark matter fraction in the central regions. While the TNG simulations are calibrated against the observed stellar mass function of galaxies, this calibration also assumes a Chabrier IMF, which does not rectify the issue of galaxy mass underestimation. In \citetalias{Wang2023} of our series of studies, we combined gravitational lensing observations with dynamical observations to reveal that the stellar mass estimates for central galaxies in groups and clusters, based on the Chabrier IMF, may be underestimated by approximately a factor of three. Furthermore, in \citet{Lu2023b} (Paper V), we found that the mass-to-light ratio of galaxies indeed aligns more closely with the predictions of a bottom-heavy model at the high-velocity-dispersion end.

It should be emphasized that our intention is not to assert that alterations in the IMF will necessarily result in a perfect agreement between observations and simulations. Instead, we aim to underscore the fixed IMF as a potentially significant factor contributing to disparities. In simulations, the recovery of the stellar mass function is often used as a benchmark for assessing simulation quality. However, when comparing with observations, simulations typically assume a fixed Chabrier IMF to compute the observed stellar mass function. This assumption makes it challenging to identify issues arising from a fixed IMF in such comparisons. Nevertheless, there have been simulations exploring the impact of variable IMFs on numerical simulation outcomes. For instance, \citet{Barber2018} introduced a variable IMF based on the initial interstellar medium (ISM) pressure and conducted simulations with fluctuating nucleosynthetic yields, star formation laws, and stellar feedback, all aligned with their locally varying IMFs. They concluded that a bottom-heavy IMF can effectively increase the stellar mass by introducing an excess of (dim) dwarf stars. Moreover, their research showed that conducting simulations with a variable IMF and making corresponding adjustments not only reproduces the observed correlation between mass-to-light ratio excess and central stellar velocity dispersion as reported in \citet{MLE}, but also achieves excellent agreement with the observational diagnostics initially employed to calibrate the subgrid feedback physics in the reference EAGLE model.

For dark halos less than $10^{10}\ {\rm M_\odot}$, the dark matter fraction in simulations is relatively close to the observed one. However, the stellar density profiles and dark matter density profiles in simulations are still steeper than observations (refer to the first row of Fig.~\ref{stkSlope}). \citet{Lovell2018} showed that in TNG, the density profile of dark matter at 5-10 kpc might be affected by the adiabatic contraction of stars \citep{Blumenthal1986, Gnedin2004}, making it steeper than results from pure dark matter simulations. In this work, we found that the stellar density profiles of TNG100 and TNG50 in the low-mass range are significantly steeper than in MaNGA galaxies, implying that the dark matter distribution in TNG simulations would be affected by a stronger contraction of baryonic matter. This may be one of the sources of discrepancies between observations and simulations.

In the past, dark halos with $M_*$ less than $10^9\ {\rm M_\odot}$ have been noted to have very flat dark matter halo density profiles, even flatter than the $\overline{\gamma}_{_{\rm DM}}=1$ prediction from the cold dark matter model \citep{Small_Challenge}. Some studies suggest that this might be related to feedback from star formation, which can rapidly expel gas from the inner dark halo, thereby reducing the gravitational potential in the center of the halo and leading to a more flattened halo density profile in the center \citep[e.g.][]{Read2005, Pontzen2012, Bentez2018, Bose2019}. Other studies propose that self-interactions of dark matter might flatten dark matter density profiles \citep{Kaplinghat2016, Tulin2018, Robertson2019, Bondarenko2021, Andrade2022, Eckert2022}. The galaxy masses involved in our study are approximately an order of magnitude higher than in these studies, so whether baryonic feedback still plays a significant role in these larger mass galaxies requires more detailed studies from future numerical simulations.

%It's worth noting that in the work of \citet{Li2019}, the total density slope obtained from the earlier MaNGA data release, whether averaged within $R_e$ or within a fixed physical radius, was also lower than that obtained from numerical simulations. In that study, comparisons were made with different hydrodynamical simulation projects, the Eagle \citep{XXXX} simulation, the Illustris Project \citep{}, and Illustris TNG Project. Despite issues with velocity dispersion measurements in the low-mass end of early MaNGA data and the impact of limited spatial resolution in numerical predictions, these issues do not qualitatively affect the results at the high-mass end.

\subsection{Sample selection}
\label{sec:Qual Distributin}
In Fig.~\ref{Compare}, we also show that the total density slope derived from \citetalias{Wang2023} by which included galaxies with a quality score of ${\rm Qual}\geq0$, whereas the results presented in this paper were based on galaxies with ${\rm Qual}>0$. If we include ${\rm Qual}=0$ galaxies in our analysis, we also predict a lower stacked $\overline{\gamma}_{_{\rm T}}$ for this mass range, consistent with the findings of \citetalias{Wang2023}.  We excluded galaxies with ${\rm Qual}=0$ in our study because of the large scatters in the total mass density slopes predicted by different mass models, even though these galaxies have a converged total mass estimate from dynamical modeling. However, \citetalias{Wang2023} only used the total mass within a certain radius from dynamical modeling, which should not be affected by the scatter of density slope from different dynamical modeling models.

\begin{figure*}
    \centering
    \includegraphics[width=2\columnwidth]{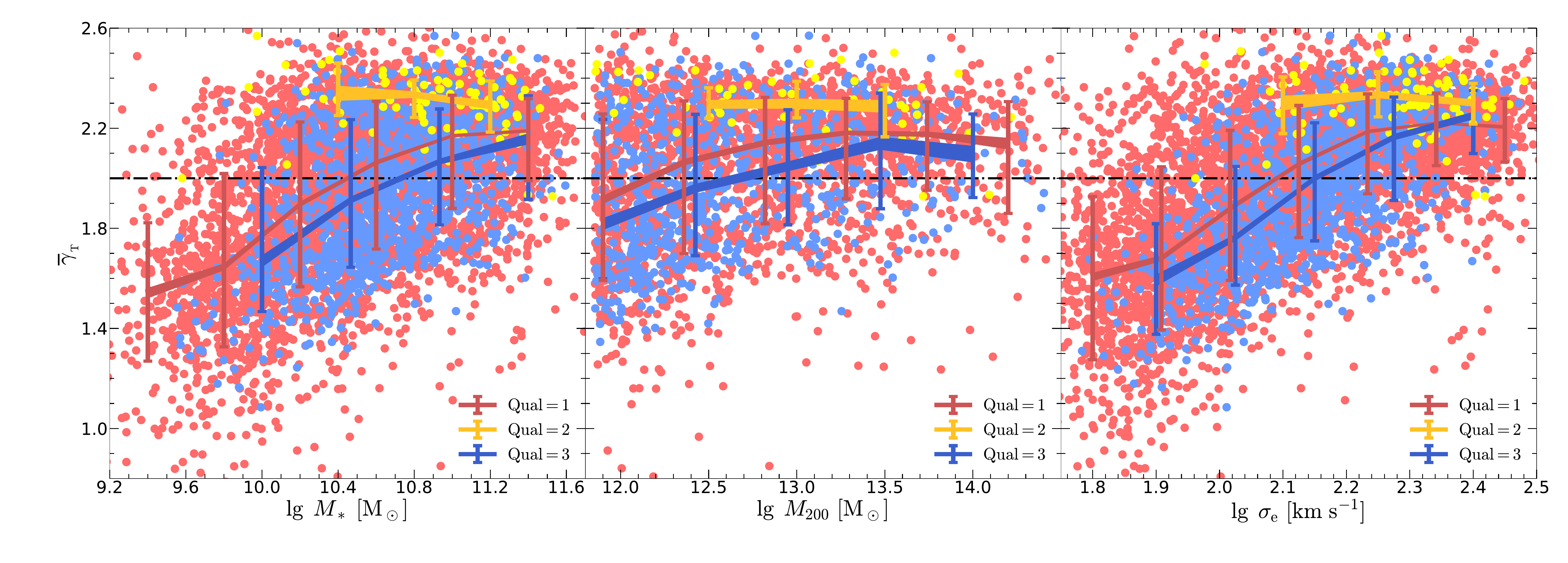}
    \caption{The $\rm Qual$ distribution of MaNGA galaxies in $\overline{\gamma}_{_{\rm T}}$-$M_*$, $\overline{\gamma}_{_{\rm T}}$-$M_{200}$ and $\overline{\gamma}_{_{\rm T}}$-$\sigma_{\rm e}$ relation. Three different colored scatter points represent $\rm Qual =1, 2, 3$ groups, respectively. Their median values are indicated by solid lines with the line width represents the stand error of the median. The error bars enclose the 1-$\sigma$ scatters. The slope value of the isothermal profile is shown with a black horizontal dash-dot line.}
    \label{Qual}
\end{figure*}

Fig.~\ref{Qual} shows the distribution of our galaxy samples with different Qual in the
$\overline{\gamma}_{_{\rm T}}$-$M_*$, $\overline{\gamma}_{_{\rm T}}$-$\sigma_{\rm e}$ and $\overline{\gamma}_{_{\rm T}}$-$M_{200}$ diagrams, excluding samples with ${\rm Qual}=-1$ since they are all untrustworthy and ${\rm Qual}=0$ since their density slopes are not reliable. Compared to the full sample, galaxies for different fitting quality groups exhibit distinct distributions. In most mass ranges, galaxies with ${\rm Qual}=3$ exhibit $\overline{\gamma}_{_{\rm T}}$ values that are 0.1 lower than galaxies with ${\rm Qual}=1$, while galaxies with ${\rm Qual}=2$ typically have $\overline{\gamma}_{_{\rm T}}$ values nearly all above 2. Although we do not show ${\rm Qual}=0$ galaxies, they have relatively shallower density slope than that of ${\rm Qual}\geq1$ galaxies. The reasons for the systematic differences in the matter distribution among galaxies with different modeling qualities need further investigation.

\section{Summary and conclusion}
\label{summary}
In this paper, we have investigated the distribution of galaxy mass by analyzing stellar kinematics modeling data from the MaNGA-DynPop project. Our study centers on approximately 6000 nearby galaxies, for which we employ reliable total density slopes obtained from the modeling work of \citetalias{Zhu2023a}. Furthermore, we explore the decomposed density profiles for both stellar and dark matter components of these galaxies. Our analysis has led to the following key findings:

\begin{enumerate}
    \item \citetalias{Zhu2023b} presents the $\overline{\gamma}_{_{\rm T}}$-$\sigma_{\rm e}$ relationships derived from the analysis of these 6000 galaxies. The updated scaling relation reveals that, for galaxies with high $\sigma_{\rm e}$ values, the average mass-weighted total density slope ($\overline{\gamma}_{_{\rm T}}$) remains relatively constant at 2.2. However, as $\sigma_{\rm e}$ decreases below a specific threshold, the galaxy density profile becomes flatter. For the full sample, this threshold is approximately 189 km/s, while for ETGs, it is approximately 150 km/s. Here, we have also observed a strong resemblance between the trends in the variation of $\overline{\gamma}_{_{\rm T}}$ and the variation of $\overline{\gamma}_*$. This similarity appears to be attributed to the dominance of stellar matter within $R_{\rm e}$ of MaNGA galaxies. 

    \item We have compared our findings with central galaxies from TNG50 and TNG100 simulations. Our analysis reveals that the current magneto-hydrodynamic cosmological simulations do not yet accurately replicate the observed galaxy's central mass distribution. In particular, both $f_{\rm DM}(<R_{\rm e})$ and $\overline{\gamma}_*$ tend to be overestimated in the simulations. Furthermore, the simulations fall short in reproducing the relationship between $\overline{\gamma}_{_{\rm T}}$ and galaxy's properties such as $M_*$, $M_{\rm 200}$, and $\sigma_{\rm e}$. We attribute these discrepancies primarily to the simulations' tendency to over-predict the dark matter fraction, which could be linked to questionable assumptions like a constant IMF, excessive adiabatic contraction effects and feedback implementations.

    \item We have observed that within a specific $\sigma_{\rm e}$ range, an increase in the stellar density slope corresponds to a higher total density slope. In this context, we have not observed dark matter compensating for the steepness of the stellar density profile to restore the density profile to a isothermal one ($\overline{\gamma}_{_{\rm T}}=2$). We conclude that there is no perfect conspiracy between baryonic matter and dark matter.
    
    \item We have presented the stacked (the average) galaxy density profiles and calculated the changes in the stacked slopes with $M_*$, $\sigma_{\rm e}$, and $M_{\rm 200}$. We find that in each sub-sample, the stacked dark matter density profile is slightly steeper than the pure dark matter simulation prediction of $r^{-1}$, which may indicate moderate adiabatic contraction in the central region of galaxy.
\end{enumerate}

Our study underscores the effectiveness of stellar dynamics modeling as a valuable tool for investigating the interaction between stellar and dark matter, thereby constraining galaxy formation theories. In the future, we plan to further our investigations by integrating stellar dynamics techniques with both strong and weak gravitational lensing data to understand galaxy formation and evolution.

\section*{Acknowledgements}
We are grateful to the referee Dr. Alessandro Sonnenfeld for constructive comments which substantially improved the quality of the paper. We acknowledge the support by National Key R\&D Program of China No. 2022YFF0503403, the support of National Nature Science Foundation of China (Nos 11988101,12022306), the support from the Ministry of Science and Technology of China (Nos. 2020SKA0110100),  the science research grants from the China Manned Space Project (Nos. CMS-CSST-2021-B01,CMS-CSST-2021-A01), CAS Project for Young Scientists in Basic Research (No. YSBR-062), and the support from K.C.Wong Education Foundation.  
% Entry for the table of contents, for this guide only

%%%%%%%%%%%%%%%%%%%%%%%%%%%%%%%%%%%%%%%%%%%%%%%%%%

\section*{Data Availability}

% \michele{Make sure you complete this before submitting, or the paper will be sent back to you.} 
%The inclusion of a Data Availability Statement is a requirement for articles published in MNRAS. Data Availability Statements provide a standardised format for readers to understand the availability of data underlying the research results described in the article. The statement may refer to original data generated in the course of the study or to third-party data analysed in the article. The statement should describe and provide means of access, where possible, by linking to the data or providing the required accession numbers for the relevant databases or DOIs.
The catalogues of dynamical quantities and the stellar population properties are publicly available on the website of MaNGA DynPop (\url{https://manga-dynpop.github.io}). The data from the IllustrisTNG simulations used in this study is publicly available on the website \url{https://www.tng-project.org/data/}. All the plotting code used and the resulting figures are available at \url{https://github.com/Shubo143/MaNGADensitySlope.git}.



\bibliographystyle{mnras}
\bibliography{ref} % if your bibtex file is called example.bib

%%%%%%%%%%%%%%%%%%%%%%%%%%%%%%%%%%%%%%%%%%%%%%%%%%

%%%%%%%%%%%%%%%%% APPENDICES %%%%%%%%%%%%%%%%%%%%%
\appendix
\section{Different behaviour of stacked stellar density slope}
\label{Appx.A}
\begin{figure*}
    \centering
    \includegraphics[width=2\columnwidth]{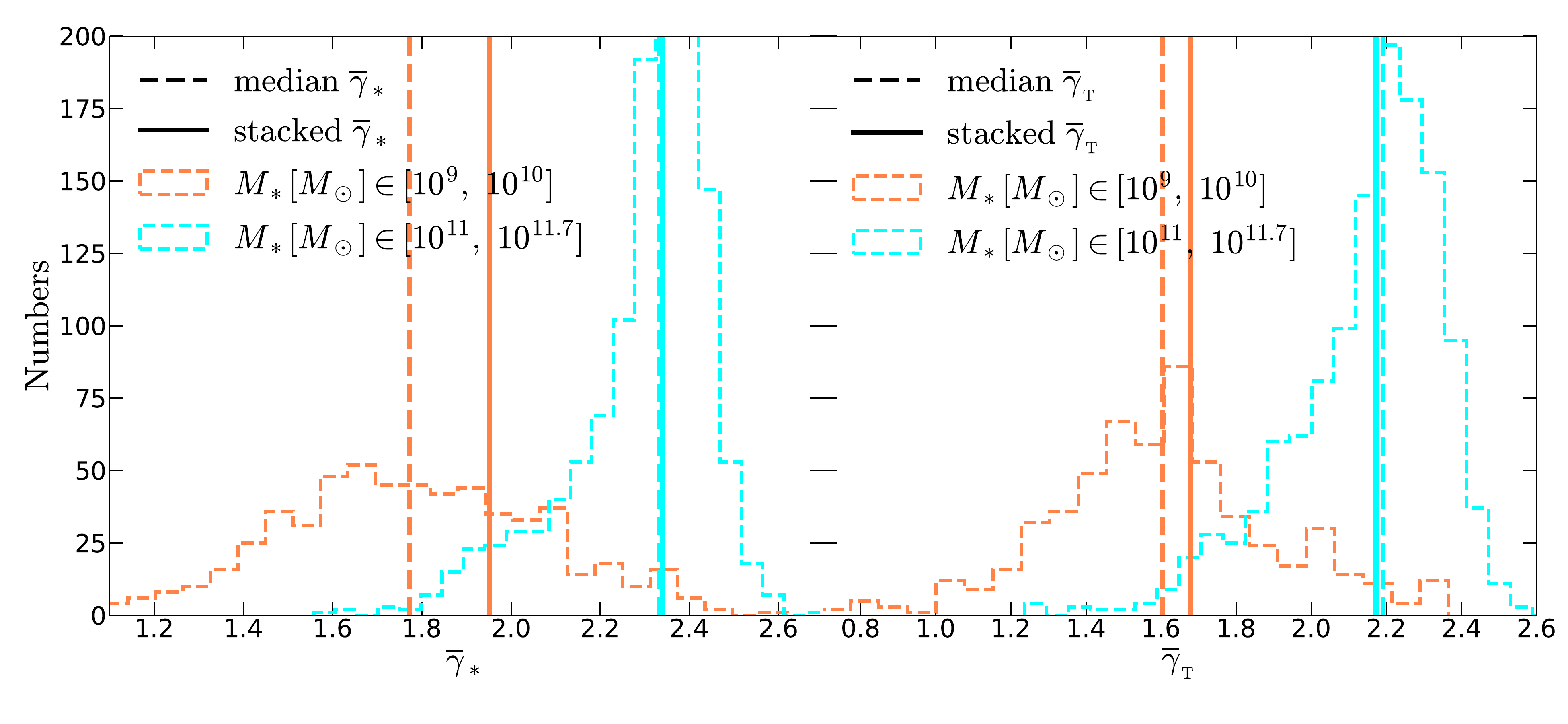}
    \caption{The distribution of $\overline{\gamma}_*$ (left panel) and $\overline{\gamma}_{_{\rm T}}$ (right panel) in the lowest $M_*$ bin (orange histogram) and highest $M_*$ bin (cyan histogram). The median values of individual galaxy's slope are indicated by vertical dashed lines with the same color of the histogram and the stacked results are represent by vertical solid lines. }
    \label{AppxFig1}
\end{figure*}

Taking $M_*$ binning as an example, in Fig.~\ref{AppxFig1} we show the distribution of $\overline{\gamma}_*$ (left panel) and $\overline{\gamma}_{_{\rm T}}$ (right panel) in the lowest $M_*$ bin (orange histogram) and the highest $M_*$ bin (cyan histogram), respectively. It can be observed that compared to $\overline{\gamma}_{_{\rm T}}$, $\overline{\gamma}_*$ exhibits a more scattered distribution in the low mass bin, reflecting the diversity of stellar profiles within this bin. We infer that calculating the mean values of these different-shaped profiles does not accurately represent the material distribution of galaxies within this mass bin. Thus, the stacked stellar density slope deviates from the median value of individual galaxy's stellar density slope more, reflecting the larger deviation for mean stellar density profile from the original stellar density profile at the low mass end.
%In Fig.~\ref{AppxFig1} the stacked slope presented is derived by taking the median profile as the stacked slope, not the mean profile we show in previous text. However, if we change back to the mean-stacked slope, the difference between stacked $\overline{\gamma}_*$ and the median of $\overline{\gamma}_*$ is still greater by $\sim0.15$ than $\overline{\gamma}_{_{\rm T}}$ and the median of $\overline{\gamma}_{_{\rm T}}$ for the low mass bin.

%%%%%%%%%%%%%%%%%%%%%%%%%%%%%%%%%%%%%%%%%%%%%%%%%%

% Don't change these lines
\bsp	% typesetting comment
\label{lastpage}
\end{document}